\documentclass[prd,aps,10pt,twocolumn,showpacs,reprint,nofootinbib]{revtex4}
 
\usepackage[dvips]{graphicx}
\usepackage{amssymb}
\usepackage{filecontents}
\usepackage{amsmath}
\usepackage{color}
\usepackage{float}
\usepackage{textcomp}
\usepackage{bm}

\RequirePackage[colorlinks,citecolor=blue,urlcolor=blue,linkcolor=blue]{hyperref}
\usepackage{yfonts}
\usepackage{amsmath,esint}

\begin{document}
\title{Constraining  energy-momentum-squared gravity by binary pulsar observations} 
\author{Elham \surname{Nazari}$^{1}$}
\email{elham.nazari@mail.um.ac.ir (Corresponding Author) }
\author{Mahmood \surname{Roshan}$^{1,2}$}
\email{mroshan@um.ac.ir}
\author{Ivan \surname{De Martino}$^{3}$}
\email{ivan.demartino@usal.es}

\affiliation{$^1$Department of Physics, Faculty of Science, Ferdowsi University of Mashhad, P.O. Box 1436, Mashhad, Iran \\
$^2$School of Astronomy, Institute for Research in Fundamental Sciences (IPM), P. O. Box 19395-5531, Tehran, Iran\\
$^3$Universidad de Salamanca, Departamento de Fisica Fundamental, P. de la Merced, Salamanca, Spain\\
}

\begin{abstract}
In this paper, we introduce the post-Minkowskian approximation of energy-momentum-squared gravity (EMSG). This approximation is used to study the gravitational energy flux in the context of EMSG. As an application of our results, we investigate the EMSG effect on the first time derivative of the orbital period of the binary pulsars. Utilizing this post-Keplerian parameter, the free parameter of the EMSG theory, $f_0'$,  is estimated for six known binary pulsars.
Taking the binaries that have the most accurate observations, it turns out that $-6\times 10^{-37}\text{m}\,\text{s}^2\,\text{kg}^{-1}<f_0'<+10^{-36}\text{m}\,\text{s}^2\,\text{kg}^{-1}$. This bound is in agreement with the precedent studies.
\end{abstract}
\maketitle

\section{Introduction}
There are several motivations behind attempts for generalizing the standard gravitational theory, namely the general theory of relativity (GR). These motivations come from both pure theoretical and also observational issues. For example, in the past decades, the dark energy enigma has been one of the main motivations to construct new gravity theories \cite{Clifton:2011jh}. An older motivation belongs to attempts for answering the dark matter problem without invoking any new particle \cite{Milgrom:1983ca,salucci2021einstein,de2020dark}. This story began by modifying the Newtonian dynamics. However, currently, there are covariant relativistic theories that in principle are significantly different from GR and claim at replacing dark matter by introducing corrections to  GR, for example, see nonlocal gravity \cite{Hehl:2008eu} and the scalar-tensor-vector-gravity theory \cite{Moffat:2005si}. 

On the other hand, changing GR is not a simple task and there are serious restrictions on the deviations from GR. The recent detection of gravitational waves (GWs), already anticipated by GR, has provided a strong tool to rule out or constrain some modified gravity theories. It is necessary to mention that modified gravity theories mostly bring new gravitational fields in addition to the metric. This means that the dynamical/gravitational degrees of freedom is increased. Consequently, the equations governing the propagation of GWs are different, in principle, from GR. More importantly, the propagation speed of GWs can be different from that of the light speed. However, we know that the event GW170817 for GWs from a binary neutron star system \cite{Abbott_2017,PhysRevLett.119.161101}, proved that at least at redshift $z\simeq 0$, the fractional relative difference in propagation speeds of GWs ($c_W$) and light ($c$) is less than $10^{-15}$. Therefore, modified theories that are incapable of satisfying $c_W=c$, are ruled out. For example, one may mention the covariant Galileon that has been ruled out by this implication \cite{Ezquiaga:2017ekz}. There are other kinds of modified gravity theories which are still viable, e.g., see \cite{boran2018gw170817}.

Another pertinent test bed for modified theories of gravity is studying the post-Keplerian parameters associated with binary pulsars \cite{de2013testing,de2011quadrupolar,de2015probing,jimenez2016evading}. These parameters provide an indirect evidence of GW emission. Historically, the time derivative of the orbital period in the relativistic binary pulsar B1913+16 has been studied in \cite{hulse1975discovery}.  
In this paper, we investigate the flux of energy due to the GW emissions from the binary systems, and consequently we obtain the first time derivative of the orbital period of several binary pulsars in the context of energy-momentum-squared gravity (EMSG).
To do so, we introduce the post-Minkowskian (PM) approximation  of this gravitational theory based on the Landau-Lifshitz formulation of the field equations.
We apply the modern approach to the PM theory introduced in \cite{will1996gravitational,pati2000post,pati2002post}

The outline of the paper is as follows. First, in Sec. \ref{emsg}, we briefly review EMSG and its current status among other modified gravity theories.  In Sec. \ref{sec2}, we clarify the strategy of our calculations. In this section, we introduce the Landau-Lifshitz formulation of the EMSG field equations and the general expression for the flux of energy in this theory. Then, by using this reformulation, the PM approximation of EMSG is comprehensively obtained in Sec. \ref{PM approximation in EMSG}. The gravitational energy flux with EMSG correction is studied in Sec. \ref{Gravitational energy flux in EMSG}. As an application of our results, we next investigate the EMSG effect on the first time derivative of the orbital period of the binary pulsar in Sec. \ref{Application to the binary pulsars}. We use this post-Keplerian parameter for six known binary pulsars to constrain one of the free parameters of the EMSG in this section. Finally, in Sec. \ref{Conclusion}, our results are discussed.

\section{Standard formulation of EMSG}
\label{emsg}

EMSG has been introduced to resolve the big bang singularity \cite{roshan2016energy,Arik:2013sti}. The main idea is to allow the scalar ${\boldsymbol{T}}^2=T^{\alpha\beta}T_{\alpha\beta}$ in the generic action of the theory, where $T_{\alpha\beta}$ is the energy-momentum tensor. The existence of this term induces some quadratic corrections to the Friedman equations that are reminiscent of the corrections reported in the context of loop quantum gravity, see \cite{roshan2016energy} for more detail. It has been shown in \cite{roshan2016energy} that EMSG has a bouncing solution preventing the early Universe singularity. However, recently in \cite{Barbar:2019rfn} the viability of this bouncing solution has been doubted. EMSG in the Palatini formulation has been investigated in \cite{Nazari:2020gnu}. It has been shown that in this approach there are viable bouncing solutions. It is important to mention that interest in EMSG, has not been limited to the early Universe and bouncing solutions. For instance, this model has been used to manipulate the cosmic microwave background quadrupole temperature fluctuation \cite{Akarsu:2020vii}. Furthermore, this model explains the accelerated expansion of the Universe without the cosmological constant. In other words, the normal matter would be enough to speed up the cosmic expansion, see \cite{Akarsu:2017ohj} and references therein.  For the phase space view of EMSG, we refer the reader to \cite{bahamonde2019dynamical}. For some other papers on the cosmological and astrophysical issues in EMSG see \cite{shahidi2021non,moraes2018nonexotic,faria2019low,sharif2021viable}.
Although the original version of EMSG does not raise significant corrections to GR at low energy scales, the current versions are necessary to be consistent with the classical tests of gravity in the solar system \cite{Nazari2021bending}.

Now let us discuss the action and the corresponding field equations of EMSG. The general action of this theory is given by 
\begin{align}\label{EMSG-action}
S=\int \sqrt{-g}\Big(\frac{1}{2k}R+f({\boldsymbol{T}}^2)\Big)d^4x+S_{M},
\end{align}
where $g$ is the determinant of the metric $g^{\alpha\beta}$, and $k=8\pi G/c^4$. Here, as usual, $R$ is the Ricci scalar, and $S_M$  is the matter action. Of course, this is not the most general action of the theory. More specifically, one may take a function of $R$, namely $f(R)$. This branch of modified gravity theories has been widely investigated in the literature. On the other hand, our focus here is on the consequences of the scalar $\boldsymbol{T}^2$. Therefore, we take the standard linear contribution for the curvature part. In contrast, we take an arbitrary function for $\boldsymbol{T}^2$.
We consider that $f({\boldsymbol{T}}^2)$ is a Taylor expandable function in terms of $\boldsymbol{T}^2$ and can then be written as 
\begin{align}
f({\boldsymbol{T}}^2)\simeq  f({\boldsymbol{T}}_0^2)+f'({\boldsymbol{T}}_0^2)\big({\boldsymbol{T}}^2-{\boldsymbol{T}}_0^2\big)+\cdots ,
\end{align}
in which the prime stands for the differentiation with respect to ${\boldsymbol{T}}^2$. Here, the suffix ``$_0$" returns to the ${\boldsymbol{T}}^2$ value in the background. On the other hand, since our focus is on the gravitational flux of energy in EMSG, we omit the dark energy contribution and assume that there is no matter field in the background. Therefore, the background is described by the flat-Minkowski metric. Then, we have ${\boldsymbol{T}}_0^2=0$ and consequently $f({\boldsymbol{T}}_0^2)= 0$.
So, the above relation reduces to
\begin{align}\label{f(T^2)}
 f({\boldsymbol{T}}^2)\simeq f'_0{\boldsymbol{T}}^2+\cdots .
\end{align}
Here, $f'_0=\frac{df}{d {\boldsymbol{T}}^2}\lvert_{{\boldsymbol{T}}^2=0}$ \footnote{In this paper, $f'_0$ is equivalent to $-\eta/2k$ in \cite{roshan2016energy}.} and regarding Eq. \eqref{EMSG-action} its SI unit is m s$^{2}$ kg$^{-1}$.
Inserting Eq. \eqref{f(T^2)} within the general form of the EMSG action \eqref{EMSG-action}, and varying each term with respect to the metric, one can obtain the field equations of EMSG as \footnote{For a more detailed discussion, we refer the reader to \cite{roshan2016energy}.}
\begin{align}\label{G}
G_{\mu\nu}=k T^{\text{eff}}_{\mu\nu},
\end{align}
where $G_{\mu\nu}=R_{\mu\nu}-\frac{1}{2}g_{\mu\nu}R$ is the Einstein tensor and 
\begin{align}\label{T_eff}
T^{\text{eff}}_{\mu\nu}=T_{\mu\nu}+f'_0\Big( g_{\mu\nu}{\boldsymbol{T}}^2-4T^\sigma_\mu T_{\nu\sigma}-4{\boldsymbol{\Psi}}_{\mu\nu}\Big),
\end{align}
is the effective conserved energy-momentum tensor:
\begin{align}\label{nabla_T_eff}
\nabla_{\mu}T_{\text{eff}}^{\mu\nu}=0.
\end{align} 
In the relation \eqref{T_eff}, the tensor ${\boldsymbol{\Psi}}_{\mu\nu}$ is given by   
\begin{align}
{\boldsymbol{\Psi}}_{\mu\nu}=-L_m\big(T_{\mu\nu}-\frac{1}{2}Tg_{\mu\nu}\big)-\frac{1}{2}TT_{\mu\nu}-2T^{\alpha\beta}\frac{\partial^2 L_m}{\partial g^{\alpha\beta}\partial g^{\mu\nu}},
\end{align}
where $L_m$ is the matter Lagrangian density and $T=g^{\alpha\beta}T_{\alpha\beta}$. To indicate the EMSG portion of the effective energy-momentum tensor and separate it from the standard part, we call the second part of Eq. \eqref{T_eff}  
\begin{align}\label{T_EMSG}
T_{\mu\nu}^{\text{\tiny EMS}}=f'_0\Big( g_{\mu\nu}{\boldsymbol{T}}^2-4T^\sigma_\mu T_{\nu\sigma}-4{\boldsymbol{\Psi}}_{\mu\nu}\Big).
\end{align}
Given the definition of ${\boldsymbol{\Psi}}_{\mu\nu}$, for the different Lagrangian densities describing a perfect fluid \cite{schutz1970perfect,brown1993action}, $T_{\mu\nu}^{\text{\tiny EMS}}$ and consequently the field equations \eqref{G} would be different in this theory. 
In fact, using different Lagrangian densities for the perfect fluid produces different field equations which indeed describe two distinct theories in the EMSG framework.
This is also the case in the $f(R)$ theories of gravity \cite{faraoni2009lagrangian,bertolami2008nonminimal}.
In the following calculations, in order to follow the literature on the EMSG theory, we use the Lagrangian density $L_{\text{m}}=p$ where $p$ is the pressure to describe a perfect fluid. We leave a supplementary study on the difference between Lagrangian densities in this theory for a future work.

We should mention that in addition to the effective energy-momentum conservation \eqref{nabla_T_eff}, we consider that the conservation of the rest-mass density $\rho$ is established in EMSG, i.e., $\nabla_{\alpha}\big(\rho u^{\alpha}\big)=0$. Here, $u^{\alpha}=\gamma(c,\boldsymbol{v})$ is the four-velocity field where $\boldsymbol{v}$ is the three-velocity field and $\gamma=u^0/c$. 
This relation is simplified so that the rescaled mass density $\rho^*=\sqrt{-g}\gamma \rho$ satisfies the following relation \cite{poisson2014gravity}: 
\begin{align}\label{rho}
\partial_t \rho^*+\partial_j(\rho^* v^j)=0.
\end{align}
In fact, $\rho^*$ is the rescaled mass density that satisfies the continuity equation.
In the following calculations, we use this useful definition of the mass density.

\section{ THE STRATEGY OF CALCULATIONS}\label{sec2}

To solve the modified version of the Einstein field equations and study the gravitational flux of energy in EMSG, we apply the PM theory. In fact, this theory provides an excellent framework in which we can approximately study the GW signals up to the required degree of accuracy. In the following, we first introduce the PM theory in EMSG and next investigate the gravitational flux of energy in this alternative theory of gravity.

In the PM approximation, it is assumed that the gravitational field is weak. As we know, this weak field limit is equivalent to the slow-motion condition for the gravitationally bound system. Since our goal is to study the bound systems, we apply these two conditions throughout our calculation.

The modern approach to the PM approximation of the gravitational field is based on the Landau-Lifshitz formulation of the Einstein field equations. Here, we apply the modern approach and introduce the Landau-Lifshitz formulation of EMSG.

\subsection{Landau-Lifshitz formulation of EMSG}

To obtain the Landau-Lifshitz formulation of the EMSG field equations,
we utilize the similar procedure introduced in \cite{poisson2014gravity}. The following identity is the main relation in this approach:
\begin{align}\label{H}
\partial_{\mu\nu}H^{\alpha\mu\beta\nu}=2(-g)G^{\alpha\beta}+\frac{16\pi G}{c^4}(-g)t_{\text{LL}}^{\alpha\beta},
\end{align}
where $H^{\alpha\mu\beta\nu}$ is the pseudotensor constructed from the gothic metric $\mathfrak{g}^{\alpha\beta}=\sqrt{-g}g^{\alpha\beta}$ as follows
\begin{align}
H^{\alpha\mu\beta\nu}=\mathfrak{g}^{\alpha\beta}\mathfrak{g}^{\mu\nu}-\mathfrak{g}^{\alpha\nu}\mathfrak{g}^{\beta\mu},
\end{align}
and it has the same symmetries as the Riemann tensor. In the above identity, $(-g)t_{\text{LL}}^{\alpha\beta}$ is the Landau-Lifshitz pseudotensor given by Eq. (6.5) of \cite{poisson2014gravity}\footnote{ In Appendix. \ref{app_1}, for the sake of convenience, the definition of this pseudotensor is also given by Eq. \eqref{tLL}.}. It should be noted that the identity \eqref{H} is valid for every spacetime.
Inserting the modified EMSG field equations, i.e., Eq. \eqref{G}, within Eq. \eqref{H}, we obtain
\begin{align}\label{Hf}
\partial_{\mu\nu}H^{\alpha\mu\beta\nu}=\frac{16\pi G}{c^4}(-g)\Big(T_{\text{eff}}^{\alpha\beta}+t_{\text{LL}}^{\alpha\beta}\Big).
\end{align}
This relation is the nontensorial form of the EMSG field equations in the Landau-Lifshitz formalism. 
In the next step toward this derivation, we consider that $\mathfrak{g}^{\alpha\beta}=\eta^{\alpha\beta}-h^{\alpha\beta}$.  Here, $h^{\alpha\beta}$ is a gravitational potential and $\eta^{\alpha\beta}$ is the Minkowski metric. Furthermore, we use the harmonic gauge condition $\partial_{\beta}h^{\alpha\beta}=0$. Applying this new definition of $\mathfrak{g}^{\alpha\beta}$ and imposing the harmonic gauge condition in Eq. \eqref{Hf}, after some manipulations, one can arrive at
\begin{align}\label{wave_eq}
\square h^{\alpha\beta}=-\frac{16\pi G}{c^4}\tau^{\alpha\beta}_{\text{eff}},
\end{align}
where $\square=\eta^{\alpha\beta}\partial_{\alpha\beta}$ is the Minkowski wave operator and 
\begin{align}\label{tau}
\tau^{\alpha\beta}_{\text{eff}}=(-g)\Big(T^{\alpha\beta}_{\text{eff}}+t_{\text{LL}}^{\alpha\beta}+t_{\text{H}}^{\alpha\beta}\Big),
\end{align}
is the effective energy-momentum pseudotensor. Here, $(-g)t_{\text{H}}^{\alpha\beta}$ is given by
\begin{align}\label{t_H}
(-g)t_{\text{H}}^{\alpha\beta}=\frac{c^4}{16\pi G}\big(\partial_\mu h^{\alpha\nu}\partial_\nu h^{\beta\mu}-h^{\mu\nu}\partial_{\mu\nu}h^{\alpha\beta}\big).
\end{align}

The wave equation \eqref{wave_eq} is the cornerstone to derive the PM approximation to the EMSG field equations.
By applying the systematic method in the PM theory, we can approximately solve the highly nonlinear wave equation \eqref{wave_eq} in the wave zone where the radiation effects are significant.   
As expected, by dropping $T_{\mu\nu}^{\text{\tiny EMS}}$ in this relation, the standard Landau-Lifshitz formulation of the Einstein field equations in GR is recovered, cf. Eq. (6.51) in \cite{poisson2014gravity}. In this case, the general form of  the solution of the wave equation is comprehensively introduced in chapter 6 in \cite{poisson2014gravity}.
Regarding the mathematical similarity of Eq. \eqref{wave_eq} and that of in GR, we can use the same general from of the solutions here. Therefore, we drop the relevant calculations. Instead, the general form of the solutions are presented in Appendix. \ref{app_1}.  

As the final point in this subsection, we should mention that the pseudotensor $(-g)t_{\text{H}}^{\alpha\beta}$ satisfies the following identity:  $\partial_{\beta}\big((-g)t_{\text{H}}^{\alpha\beta}\big)=0$.
Keeping this fact in mind and also using the harmonic gauge condition, we deduce from Eq. \eqref{wave_eq} that 
\begin{align}\label{cons_eq}
\partial_{\mu}\Big((-g)(T^{\mu\nu}_{\text{eff}}+t_{\text{LL}}^{\mu\nu})\Big)=0.
\end{align}
In fact, this is a statement of the conservation equations in this formalism. The wave equation \eqref{wave_eq} and the conservation equations \eqref{cons_eq} are the main relations to find the PM approximation to the EMSG field equations.

\subsection{Flux of the energy}\label{flux of the energy}

Let us introduce the energy-balance equation in the Landau-Lifshitz formalism of EMSG. As we know, by studying this equation, one can obtain the relation between the time dependent physical properties of the systems and the amount of energy carried away by the GW emissions.
To do so, in a similar fashion to the method introduced in the Landau-Lifshitz formalism of GR,
we first introduce the total four-vector momentum as
\begin{align}
P^{\alpha}[V]=\frac{1}{c}\int_{V}(-g)\big(T^{\alpha 0}_\text{eff}+t_{\text{LL}}^{\alpha 0}\big)d^3 x.
\end{align}
Here, $V$ is the volume of the three-dimensional region where $P^{\alpha}[V]$ is defined.
The time and space components of this four-vector momentum represent the total energy $P^0[V]=c^{-1}E[V]$ and three-vector momentum $P^j[V]$ associated with $V$, respectively. 
It is also considered that there is no matter field on the boundary of this region. In other words, $V$ is large enough so that the surface of $V$, indicated by $S$, does not intersect the matter distribution. Bearing this fact in mind and also using the conservation equations \eqref{cons_eq}, one can arrive at
\begin{align}\label{dot_P}
\dot{P}^{\alpha}[V]=-\oint_{S}(-g)t_{\text{LL}}^{\alpha k}d S_k,
\end{align}
for the conservation statement of the total four-vector momentum. Here, the overdot stands for the derivative with respect to time $t$.
At the first glance, this relation reveals that not only the standard energy-momentum tensor but also the EMSG portion $T_{\mu\nu}^{\text{\tiny EMS}}$ have no contribution in this conservation statement and the mathematical form of this equation is the same as in GR \cite{poisson2014gravity}. However, it should be mentioned that the Landau-Lifshitz pseudotensor is not necessarily similar in these two theories. In the following section, we obtain this pseudotensor in EMSG and illustrate the differences.

As we aim to study the rate at which the energy is removed from the GW sources, we focus our attention on the time component of Eq. \eqref{dot_P}. Then for $\alpha=0$, we have
\begin{align}\label{dot_E}
& \dot{E}[V]=-c\oint_{S}(-g)t_{\text{LL}}^{0 k}d S_k.
\end{align}
Here, 
\begin{align}
E[V]=\int_{V}(-g)\big(T^{00}_\text{eff}+t^{00}_{\text{LL}}\big)d^3x,
\end{align}
is the total energy of the system.
Regarding Eq. \eqref{dot_E}, one can define the energy-balance equation in EMSG as 
\begin{align}\label{energy-balance_equation}
\frac{d E}{d t}=-\mathcal{P},
\end{align}
where
\begin{align}\label{eq01}
\mathcal{P}=c\oint_{S}(-g)t_{\text{LL}}^{0k}d S_k,
\end{align}
is the flux of energy across the surface $S$.

The next step toward studying the energy-balance equation is to find the Landau-Lifshitz pseudotensor in terms of the gravitational potential $h^{\alpha\beta}$. 
It is shown in the shortwave approximation where the characteristic wavelength of the GW signals $\lambda_c$ is much smaller than the distance to the center of the mass of the GW system $r=\vert \mathbf{x}\vert$, i.e., where $\lambda_c/r\ll 1$, one can simplify the definition \eqref{tLL} as \footnote{ To see more detailed discussions on this issue, we refer the reader to Sec. 12.2.3 of \cite{poisson2014gravity}.}
\begin{align}\label{t_LL}
(-g)t_{\text{LL}}^{\alpha\beta}=\frac{c^2}{32 \pi G}\Big(\partial_\tau h_{\text{TT}}^{jk}\partial_\tau h^{\text{TT}}_{jk}\Big)k^{\alpha}k^{\beta},
\end{align}
where $\partial_\tau$ shows the derivative with respect to retarded time $\tau=t-r/c$ and four vector $k^{\alpha}$ is defined by $k^{\alpha}=(1,\boldsymbol{n})$. Here  $h^{jk}_{\text{TT}}=\Big(q_l^jq_m^k-\frac{1}{2}q^{jk}q_{lm}\big)h^{lm}$ represents the transverse-tracefree (T-T) portion of the gravitational potential $h^{jk}$ in which $q_{jk}=\delta_{jk}-n_jn_k$ and $n_j=\partial_{j}r$. As seen, the mathematical appearance of Eqs. \eqref{energy-balance_equation} and \eqref{eq01} is similar to the GR one. However, $h_{\text{TT}}^{jk}$ can be basically different in EMSG. So, it is not trivial how GW systems will behave in the context of this modified theory of gravity.

To sum up, in order to obtain $(-g)t_{\text{LL}}^{\alpha\beta}$ and consequently the flux of energy in EMSG, we should find T-T part of the gravitational potential in this theory. So, we turn to solve the EMSG field equations \eqref{wave_eq}. In the next section, we attempt to solve these equations by utilizing the PM approximation and obtain $h_{\text{TT}}^{jk}$ to the leading order in the EMSG theory. 
Those readers who are unfamiliar with the PM approximation, or have no interest in the detailed derivation, can directly skip to Sec. \ref{Gravitational energy flux in EMSG}. In this section, applying the results of  Sec. \ref{PM approximation in EMSG}, we study the flux of energy due to the GW emissions in EMSG.

\subsection{Crude estimation}

Before starting the main calculations, let us approximately obtain the GW field $h_{\text{TT}}^{jk}$ for a specific distribution of the matter in EMSG. We consider the famous quadrupole formula for $h_{\text{TT}}^{jk}$, i.e., $h_{\text{TT}}^{jk}=(2G/c^4 r) \partial_{\tau \tau}\mathcal{I}^{jk}$, in which 
\begin{align}
\mathcal{I}^{jk}(\tau)=\int c^{-2}\tau^{00}(\tau,\mathbf{x}')x'^jx'^kd^3x',
\end{align} 
is the quadrupole-moment tensor of the matter distribution. In the EMSG case, one may replace $\tau^{00}$ with $\tau_{\text{eff}}^{00}$. To find $h_{\text{TT}}^{jk}$, we should then derive the time-time component of the effective energy-momentum pseudotensor \eqref{tau}. 
To do so, we choose a perfect fluid for describing the matter distribution in the Minkowski spacetime as $T^{\alpha\beta}=\big(\rho+\epsilon/c^2+p/c^2\big)u^{\alpha}u^{\beta}+p\eta^{\alpha\beta}$.
Here,  $\epsilon$ is the proper internal energy density and $p$ is the pressure.  
We also consider that the velocity of the fluid elements is slow compared to the speed of light and the following set of conditions is established in this fluid system ($v^2/c^2\ll 1,\, p/\rho c^2\ll 1,\, \text{and}\, \epsilon/\rho c^2\ll 1$). With this in mind, regarding Eqs. \eqref{T_eff}, \eqref{tLL_harmonic}, and \eqref{t_H}, one can show that $\tau^{00}_{\text{eff}}$ up to the leading order is given by 
\begin{align}
\tau^{00}_{\text{eff}}=c^2\rho_{\text{\tiny EMS}}^{\text{eff}}+O(1),
\end{align}
where $\rho_{\text{\tiny EMS}}^{\text{eff}}=\rho\left(1+c^2f_0'\rho\right)$ represents the effective mass density in EMSG. So, we can conclude that finding the quadrupole moment 
\begin{align}
\mathcal{I}^{jk}(\tau)=\int \rho_{\text{\tiny EMS}}^{\text{eff}}(\tau,\mathbf{x}')x'^jx'^kd^3x',
\end{align}
will be the main task toward deriving $h_{\text{TT}}^{jk}$ in EMSG.  
This crude estimation clearly shows that not only does $\rho$ play a role but $\rho^2$ also has a contribution to the GW field $h_{\text{TT}}^{jk}$. On the other hand, it should be emphasized that as this estimation is based on the behavior of the perfect fluid in the flat spacetime, the matter field indeed does not respond to gravity.
In other words, enforcing Eq. \eqref{nabla_T_eff} for the matter field in the Minkowski spacetime reveals that the matter is not subjected to the gravitational interaction in this estimation.
Moreover, the Landau-Lifshitz and harmonic pseudotensors vanish here. 
Therefore, to incorporate the role of the gravity into the result, it is necessary to obtain $\tau^{00}_{\text{eff}}$ more precisely. Toward this goal, in the next section, we introduce the PM expansion of the gravitational potentials in EMSG.

As it will be shown, we also need to indicate the order of magnitude of the free parameter of this theory. To do so, let us recall the time-time component of the metric up to the Newtonian order: $g_{00}=-1+2U/c^2$ where $U$ is the gravitational potential and given by $\nabla^2 U=-4\pi G \rho$. One may deduce that this metric component will be $g_{00}=-1+2U_{\text{\tiny N}}/c^2+2f_0'U_{\text{\tiny EMS}}$ after changing $\rho$ to the effective mass density $\rho_{\text{\tiny EMS}}^{\text{eff}}$. Here, $\nabla^2 U_{\text{\tiny N}}=-4\pi G \rho$ and  $\nabla^2 U_{\text{\tiny EMS}}=-4\pi G \rho^2$ are the Poisson equations for the Newtonian and EMSG potentials $U_{\text{\tiny N}}$ and $U_{\text{\tiny EMS}}$, respectively. Considering the weak field limit and comparing the first, second, and third terms in this component of the metric reveal that $U_{\text{\tiny N}}/c^2$ and $f_0'U_{\text{\tiny EMS}}$ should be very small, i.e., $U_{\text{\tiny N}}/c^2\ll 1$ and $f_0'U_{\text{\tiny EMS}}\ll 1$.  So, we consider that $f_0'U_{\text{\tiny EMS}}$ is at most of the order of $U_{\text{\tiny N}}/c^2$ \footnote{In the following, for the sake of simplification, we drop the index ``N ".}.

\section{PM approximation in EMSG}\label{PM approximation in EMSG}

As mentioned earlier, in the context of the PM approximation, the system is restricted by two constraints: the slow-motion condition and weak-field limit, i.e., $v^2/c^2\ll 1$ and $U/c^2\ll 1$, respectively. Regarding the last paragraph in the previous subsection, we would also have $f_0'U_{\text{\tiny EMS}}\ll 1$ in the weak-field limit.  Given this point, hereafter, we treat each order of $f_0'$ as a correction of the order $c^{-2}$.
In this approximate context, the spacetime deviates slightly from the flat spacetime. It is considered that its deviation, i.e., the gravitational potential $h^{\alpha\beta}$, can be expanded in the power of the gravitational constant $G$ asymptotically \cite{poisson2014gravity}. This is known as the PM expansion of the gravitational potentials. 
Moreover, in this framework, each order $c^{-2}$ is considered as a post-Newtonian (PN) correction which indicates the order of the magnitude of each term. 
In the modern approach to the PM theory, using the PM expansion of the gravitational potentials and applying the iterative procedure, one can approximately solve the highly nonlinear field equations and obtain $h^{\alpha\beta}$ to the required PN order. The main point in the iterative procedure is that the source term of the wave equation \eqref{wave_eq} in the $n$th iterated step is constructed in the previous step, i.e., $(n-1)$th iterated step. Then the wave equation will be linear and can be solved systematically.
In the following, we take the advantages of this method to solve Eq. \eqref{wave_eq} in the wave zone.

The PN expansion of the spacetime metric is the main material required at least in the first steps of our derivation.
To construct the PN expansion of the metric, we need to have information about the PN order of the gravitational potentials.
For a survey of the PN order of $h^{\alpha\beta}$, let us estimate the leading PN order of the first term in the PM expansion of the gravitational potentials, $h^{\alpha\beta}_{(1)}=O(G)$, by using the following crude scheme.
We will obtain the complete PN expansion of $h^{\alpha\beta}$ to the required degree of accuracy in the subsequent subsections.

Let us introduce the standard and EMSG portions of the energy-momentum tensors of a specific system. 
Here, we consider that the system is a perfect fluid. So, we have
\begin{align}\label{T_PF}
T^{\alpha\beta}=\big(\rho+\frac{\epsilon}{c^2}+\frac{p}{c^2}\big)u^{\alpha}u^{\beta}+pg^{\alpha\beta},
\end{align} 
for the standard energy-momentum tensor.
Moreover, utilizing the perfect fluid description, we find
\begin{align}
\nonumber
& T^{\alpha\beta}_{\text{\tiny EMS}}=f_0'\Big(c^4\rho^2g^{\alpha\beta}+2c^2\big(\epsilon\rho g^{\alpha\beta}+\rho^2u^{\alpha}u^{\beta}\big)+\big(3p^2\\\label{T_PF_EMSG}
&+\epsilon^2\big)g^{\alpha\beta}+\big(8p\rho+4\epsilon\rho+\frac{1}{c^2}(6p^2+8p\epsilon+2\epsilon^2)\big)u^{\alpha}u^{\beta}\Big),
\end{align}
for the EMSG part of energy-momentum tensor \eqref{T_EMSG}.
To arrive at this relation, we use the normalization condition $g_{\alpha\beta}u^{\alpha}u^{\beta}=-c^2$ and assume that $L_m=p$ for a perfect fluid \cite{schutz1970perfect}.
In the simplest case, we assume that the spacetime is described by the flat Minkowski metric $\eta^{\alpha\beta}$.
So, it is obvious that leading PN order of $T^{00}$ is $O(c^2)$, $T^{0j}$ is $O(c)$, and $T^{jk}$ is $O(1)$. Also, $T^{00}_{\text{\tiny EMS}}$ is of the order $c^2$, $T^{0j}_{\text{\tiny EMS}}$ is of the order $c$, and $T^{jk}_{\text{\tiny EMS}}$ is of the order $c^2$ when $j=k$ and of the order $c^0$ when $j\neq k$.
Regarding the coefficient $c^{-4}$ in Eq. \eqref{wave_eq}, we then conclude that in EMSG, the leading PN correction of $h^{00}$ is of the order $c^{-2}$, $h^{0j}$ is of the order $c^{-3}$, and $h^{jk}$ is of the order $c^{-2}$ when $j=k$ and of the order $c^{-4}$ when $j\neq k$.
This crude estimation of the PN order of $h^{\alpha\beta}$'s components illuminates our path toward finding the correct required degree of accuracy for the PN metric.

\subsection{ Post-Minkowskian expansion of metric}\label{PM_expansion_of_metric}

Regarding the estimation of the PN order of the $h^{\alpha\beta}$ components stated before as well as utilizing Eq. \eqref{PM_metric}, we find the PM expansion of the metric components in terms of the gravitational potentials as 
\begin{subequations}
\begin{align}
\nonumber
&
 g_{00}=-1+\frac{1}{2}h^{00}-\frac{3}{8}\big(h^{00}\big)^2+\frac{1}{2}h^{kk}\big(1-\frac{1}{2}h^{00}\big)\\\label{g00}
&-\frac{1}{8}\big(h^{kk}\big)^2+O(c^{-6}),\\
\label{g0j}
& g_{0j}=-h^{0j}+O(c^{-5}),\\
\label{gjk}
& g_{ij}=\delta_{ij}\Big(1+\frac{1}{2}h^{00}\Big)+h^{ij}-\frac{1}{2}\delta_{ij}h^{kk}+O(c^{-4}).
\end{align}
\end{subequations}
These equations provide sufficient PN corrections for the metric components that we need in the following computation.
In an equivalent manner, using the general equation \eqref{PM_g}, we arrive at
\begin{align}
\label{g}
&(-g)=1+h^{00}-h^{kk}+O(c^{-4}),
\end{align}
for the PM expansion of $g$ containing the PN corrections to $O(c^{-4})$ order. 
As expected from the high order of $h^{kk}$, in this modified theory of gravity, the PM expansions of the metric and its determinant are more complicated than the GR ones.

\subsection{Gravitational potentials in the first iteration}\label{first iteration}

After choosing the matter portion of the system, we are now in the position to solve the EMSG version of the field equations in the Landau-Lifshitz formalism. 
To do so, we utilize the iteration procedure. 
We recall that although our aim is to obtain $h^{jk}$ in the wave zone, enough information about the metric in the near zone, especially in the first iteration step is required. Let us mention that, in the PM theory, the near zone is restricted to a three-dimension sphere with the radius $\mathcal{R}\approx\lambda_{\text{c}}$. 
Beyond this region is called the wave zone. Given the position of the source point in the solution of the wave equations, the gravitational potential is separated into two parts. In a similar fashion to \cite{poisson2014gravity}, we call these parts the near-zone and wave-zone portions of the gravitational potential when the source point is situated in the near and wave zones, respectively. Of course, the solution of the wave equations also depends on the position of the field point. In each case, we specify the field point position.

We now launch the iteration procedure. 
In the zeroth iterated step, we assume that $h_{\alpha\beta}^{(0)}=0$ and $g_{\alpha\beta}^{(0)}=\eta_{\alpha\beta}$.
It should be emphasized that the assumption $h_{\alpha\beta}^{(0)}=0$ is equivalent to what we have considered to derive Eq. \eqref{f(T^2)}. In fact, here, it is assumed that there is no matter field in the background.
Therefore, we have $\sqrt{-g^{(0)}}=1$ and $\gamma=1+\frac{1}{2}\frac{v^2}{c^2}+O(c^{-4})$. By considering the definition of rescaled mass density, one can simply deduce that $\rho=\big(1-\frac{1}{2}\frac{v^2}{c^2}+O(c^{-4})\big)\rho^*$ in the zeroth iteration.
 We then turn to find $h_{\alpha\beta}^{(1)}$ in the first iterated step. As we will see in the second iterated step, in order to extract the correct and required PN order for  $h_{\alpha\beta}^{(2)}$, we need to find  $h_{(1)}^{00}$ to $O(c^{-2})$, $h_{(1)}^{0j}$ to $O(c^{-3})$, and $h_{(1)}^{ij}$ up to $O(c^{-4})$ in the first iterated step. So, we should derive the sources of these gravitational potentials, i.e., the components of the effective energy-momentum tensor, as follows: $\tau^{00}_{\text{(0)eff}}$ to $O(c^{2})$, $\tau^{0j}_{\text{(0)eff}}$ to $O(c)$, $\tau^{ij}_{\text{(0)eff }}$ up to $O(1)$.
In the following, we introduce each part of $\tau^{\alpha\beta}_{\text{(0)eff}}$ separately. For the usual energy-momentum tensor, $T^{\alpha\beta}_{(0)}$, we have 
\begin{subequations}
\begin{align}\label{T(0)}
& c^{-2}T^{00}_{(0)}=\rho^*+O(c^{-2}),\\
& c^{-1}T^{0j}_{(0)}=\rho^*v^j+O(c^{-2}),\\
& T^{ij}_{(0)}=O(1).
\end{align}
Also, for the EMSG energy-momentum tensor \eqref{T_PF_EMSG} up to the similar PN order, we find that
\end{subequations}
\begin{subequations}
\begin{align}
\label{T00(0)EMS}
& c^{-2}T^{00}_{\text{\tiny(0)EMS}}=2c^2 f_0'{\rho^*}^2+O(c^{-2}),\\
\label{T0j(0)EMS}
& c^{-1}T^{0j}_{\text{\tiny(0)EMS}}=2c^2 f_0'{\rho^*}^2v^j+O(c^{-2}),\\
\label{T(0)EMSG}
&
 T^{ij}_{\text{\tiny(0)EMS}}=c^4f_0'{\rho^*}^2\delta^{ij}+O(1).
\end{align}
\end{subequations} 
Furthermore, since in the zeroth iterated step $h^{\alpha\beta}_{(0)}=0$, we arrive at $t_{(0)\text{LL}}^{\alpha\beta}=0=t_{(0)\text{H}}^{\alpha\beta}$. Hence, we obtain the main materials desired to construct the gravitational potentials in the next iteration of the gravitational field equations.
Now, we attempt to find the gravitational potentials where both the field and source points are situated in the near zone in the first iterated step, i.e., $h_{(1)\mathcal{N}}^{\alpha\beta}$.  
To do so, we insert Eqs. \eqref{T(0)}-\eqref{T(0)EMSG} into integral \eqref{hNear}.
For the time-time component of $h^{\alpha\beta}_{(1)}$, we have
\begin{align}
\label{h00_N1}
& h_{(1)\mathcal{N}}^{00}=\frac{4}{c^2}U+4f_0'U_{\text{\tiny EMS}}+O(c^{-3}),
\end{align}
in which $U$ is the Newtonian potential in terms of $\rho^*$ and $U_{\text{\tiny EMS}}$ is the EMSG gravitational potential. The Poisson integrals of $U$ and $U_{\text{\tiny EMS}}$ are given by 
\begin{subequations}
\begin{align}
\label{U1}
& U=G\int_{\mathcal{M}}\frac{{\rho^*}'}{\rvert{\boldsymbol{x}-\boldsymbol{x}'}\rvert}d^3x',\\
\label{U_EMSG}
& U_{\text{\tiny EMS}}=G\int_{\mathcal{M}}\frac{{{\rho^*}'}^2}{\rvert\boldsymbol{x}-\boldsymbol{x}'\rvert}d^3x',
\end{align}
\end{subequations}
respectively.
Here, $\mathcal{M}$ represents the three-dimension region which in fact separates the near zone from the wave zone.
Similarly, for the time-space components of $h^{\alpha\beta}_{(1)}$ in the required order,
we find that
\begin{align}
\label{h0j_N1}
& h^{0j}_{(1)\mathcal{N}}=\frac{4}{c^3}U^{j}+\frac{8}{c}f_0'U^j_{\text{\tiny EMS}}+O(c^{-4}),
\end{align}
where $U^j$ is the well-known PN vector potential
\begin{align}
U^j=G\int_{\mathcal{M}}\frac{{\rho^*}'v'^j}{\rvert{\boldsymbol{x}-\boldsymbol{x}'}\rvert}d^3x',
\end{align}
while $U^j_{\text{\tiny EMS}}$ is the new vector potential in the EMSG framework defined as
\begin{subequations}
\begin{align}
& U^j_{\text{\tiny EMS}}=G\int_{\mathcal{M}}\frac{{{\rho^*}'}^2v'^j}{\rvert{\boldsymbol{x}-\boldsymbol{x}'}\rvert}d^3x'.
\end{align}
\end{subequations}
Finally, for the space-space components, we arrive at 
\begin{align}
\label{hjk_N1}
& h^{ij}_{(1)\mathcal{N}}= 4f_0' \delta^{ij}\Big( U_{\text{\tiny EMS}} -\frac{G}{c}\frac{d}{dt}\mathfrak{M}\Big)+O(c^{-4}),
\end{align}
in which 
\begin{align}\label{D1}
\mathfrak{M}=\int_{\mathcal{M}}{{\rho^*}'}^2d^3x',
\end{align}
is a new quantity in EMSG.
It should be noted that all integrands in the above relations are a function of $t$ and ${\boldsymbol{x}'}$. 
So far, we have found the near-zone portion of the gravitational potential. To complete our derivation in this iterated step, we should also find the wave-zone part of the gravitational potential, i.e., $h_{(1)\mathcal{W}}^{\alpha\beta}$ where the source point is in the wave zone and the field point is situated in the near zone.
In this case, as mentioned before, the source term of the potential $h_{(1)}^{\alpha\beta}$ is equal to $T_{(0)\text{eff}}^{\alpha\beta}$; and $t_{(0)\text{LL}}^{\alpha\beta}$ and $t_{(0)\text{H}}^{\alpha\beta}$ have no portion. Therefore, the source term is completely constructed from the matter part in this step. 
On the other hand, the matter part of the system is completely restricted to the near zone when the slow-motion condition is established \cite{poisson2014gravity}. As we consider a gravitationally bound system, this assumption is also true throughout our derivation.
This fact reveals that the wave-zone part of the gravitational potential in the first iterated step is zero. Subsequently, we have $h_{(1)\mathcal{W}}^{\alpha\beta}=0$ and $h_{(1)}^{\alpha\beta}=h_{(1)\mathcal{N}}^{\alpha\beta}$.

To sum up, we obtain the PN expansion of the EMSG potential's components required to construct the near-zone metric in the first iterated step. Substituting Eqs. \eqref{h00_N1}, \eqref{h0j_N1}, and \eqref{hjk_N1} within Eqs. \eqref{g00}, \eqref{g0j},\eqref{gjk}, and \eqref{g}, we finally arrive at
\begin{subequations}
\begin{align}
\label{g00_1}
&g_{00}^{(1)}=-1+\frac{2U}{c^2}+8f_0'U_{\text{\tiny EMS}}+O(c^{-3}),\\
\label{g0j_1}
&g_{0j}^{(1)}=-\frac{4U^j}{c^3}-\frac{8}{c}f_0'U^j_{\text{\tiny EMS}}+O(c^{-4}),\\
\label{gij_1}
&g_{ij}^{(1)}=\Big(1+\frac{2U}{c^2}\Big)\delta_{ij}+O(c^{-4}),\\
&(-g_{(1)})=1+\frac{4U}{c^2}-8f_0'U_{\text{\tiny EMS}}+O(c^{-3}).
\end{align}
\end{subequations}
Here, we discard some extra PN orders.

As seen, the components and determinant of the near-zone metric are constructed from the usual potentials $U$ and $U^j$ as well as the EMSG potentials $U_{\text{\tiny EMS}}$ and $U^j_{\text{\tiny EMS}}$. Notice that in order to describe a relativistic system to the first PN (1\tiny PN \normalsize) order, one should find the PN corrections of $g^{00}$, $g^{0j}$, and $g^{jk}$ to $O(c^{-4})$, $O(c^{-3})$, and $O(c^{-2})$, respectively. It is obvious from the above terms, in this iterated step, the required PN orders for the time-time metric component are not completely evaluated. It means that one should continue the journey toward the next iterated step and find $g^{(2)}_{\alpha\beta}$.
It is worth noting that the odd order $c^{-3}$ that appears in the determinant and time-time component of the metric is entirely built of the EMSG term that is proportional to $ d\mathfrak{M}/dt$. In Appendix. \ref{app_2}, we simplify this term. 
However, because in this step, the required PN orders of the metric to describe a relativistic system are not built, we are not concerned about this odd-power order. In other words, after finding the metric components at least to the 1\tiny PN \normalsize order, one can truly interpret the role of the odd PN orders and examine the time-reversal invariance of solutions.

\subsection{Gravitational potentials in the second iteration}\label{Gravitational potentials in the second iteration}

In the following, we attempt to obtain the appropriate PN corrections for $\tau^{\alpha\beta}_{\text{(1)eff}}$ that are necessary to derive $h^{\alpha\beta}_{(2)}$. 
Regarding the relationship between the effective energy-momentum pseudotensor and gravitational potential components, we deduce that the source terms of the gravitational potential should be built to  $O(c^2)$ for $\tau^{00}_{\text{(1)eff}}$, $O(c)$ for $\tau^{0j}_{\text{(1)eff}}$, and $O(1)$ for $\tau^{jk}_{\text{ (1)eff}}$. This fact will be clarified in the following computations.

By inserting Eqs.  \eqref{g00_1}-\eqref{gij_1} into the normalization condition $g_{\alpha\beta}u^{\alpha} u^{\beta}=-c^2$, we find that
$\gamma^{(1)}=1+4f_0'U_{\text{\tiny EMS}}+\frac{1}{c^2}\big(\frac{1}{2}v^2+U\big)+O(c^{-3})$.
Then, using $\gamma^{(1)}$ and $g_{(1)}$, we have
$\rho^*=\Big[1+\frac{1}{c^2}\big(\frac{1}{2}v^2+3U\big)\Big]\rho+O(c^{-4})$ for the rescaled mass density. Now, by applying these parameters and the components of $g_{\alpha\beta}^{(1)}$ in Eq. \eqref{T_PF}, we arrive at
\begin{subequations}
\begin{align}
\label{T00_1}
& T^{00}_{(1)}=\rho^*c^2+O(1),\\
& T^{0j}_{(1)}=\rho^*v^jc+O(c^{-1}),\\
\label{T^jk_1}
& T^{jk}_{(1)}=p\delta^{jk}+\rho^*v^jv^k+O(c^{-2}),
\end{align}
\end{subequations}
for the contravariant components of the energy-momentum tensor in the first iterated step. We keep terms up to the desired PN order mentioned earlier. It is worth mentioning that in Eq. \eqref{T00_1}, the $O(1)$ term is entirely constructed from the EMSG term and it is given by $\rho^*c^2f_0'U_{\text{\tiny EMS}}$.

We now obtain the EMSG part of the energy-momentum tensor $T^{\alpha\beta}_{\text{\tiny (1)EMS}}$. According to the original definition of this tensor, i.e., Eq. \eqref{T_EMSG}, we first obtain ${\boldsymbol{T}}^2_{(1)}$ for the perfect fluid up to $O(1)$. This is the appropriate PN order to find the required PN expansion of $T^{\alpha\beta}_{\text{\tiny (1)EMS}}$'s components.
In this step, one can deduce  
\begin{align}
{\boldsymbol{T}}^2_{(1)}={\rho^*}^2c^4-{\rho^*}^2c^2\big(v^2-2\Pi+6U\big)+O(1).
\end{align}
Here, $\Pi= \epsilon/\rho^*$.
Then using the above relation and the PN expansion of the metric and energy-momentum tensor components in the first iterated step, we arrive at
\begin{subequations}
\begin{align}
\label{T^00_1_EMS}
& T^{00}_{\text{\tiny (1)EMS}}={\rho^*}^2c^4f_0'+O(1),\\
& T^{0j}_{\text{\tiny (1)EMS}}=2{\rho^*}^2c^3f_0'v^j+O(c^{-1}),\\
\nonumber
& T^{jk}_{\text{\tiny (1)EMS}}={\rho^*}^2c^4f_0'\Big[\delta^{jk}+\frac{1}{c^2}\Big(2v^jv^k-\big(v^2-2\Pi\\\label{T^jk_1_EMS}
&+8U\big)\delta^{jk}\Big)\Big]+O(c^{-2}),
\end{align}
\end{subequations}
after some simplifications. Here, we again truncate the PN expansion of these components to the required accuracy.
It should be mentioned that the $c^{0}$ order in $T^{00}_{\text{\tiny (1)EMS}}$ is built of the EMSG term that is proportional to $ {\rho^*}^2c^4{f_0'}^2U_{\text{\tiny EMS}}$.

The remaining source terms of the gravitational potential $h_{(2)}^{\alpha\beta}$ are $(-g^{(1)})t_{(1)\text{LL}}^{\alpha\beta}$ and $(-g^{(1)})t_{(1)\text{H}}^{\alpha\beta}$. As we have mentioned before, these pseudotensors themselves are constructed from the gravitational potential. In the first iterated step, they are built of $h_{(1)}^{\alpha\beta}$. 
We keep those PN orders in the expansion of $h_{(1)}^{\alpha\beta}$ which construct the order $c^{2}$ for $(-g^{(1)})t_{(1)\text{LL,H}}^{00}$, order $c^{1}$ for $(-g^{(1)})t_{(1)\text{LL,H}}^{0j}$, and order $c^{0}$ for $(-g^{(1)})t_{(1)\text{LL,H}}^{jk}$.
We start with the Landau-Lifshitz pseudotensor. 
After inserting Eqs. \eqref{h00_N1}, \eqref{h0j_N1}, and \eqref{hjk_N1} within the definition \eqref{tLL_harmonic}, we find 
\begin{subequations}
\begin{align}
\label{t^00_1_LL}
& (-g^{(1)})t_{(1)\text{LL}}^{00}=O(1),\\
& (-g^{(1)})t_{(1)\text{LL}}^{0j}=O(c^{-1}),\\
\label{t^jk_1_LL}
& (-g^{(1)})t_{(1)\text{LL}}^{jk}= \frac{1}{4\pi G}\bigg[\partial^jU\partial^kU-\frac{1}{2}\partial^nU\partial_nU\delta^{jk}\bigg]\\
\nonumber
&+\frac{c^4f_0'}{\pi G}\bigg[f_0'\Big(\partial^nU_{\text{\tiny EMS}}\partial_nU_{\text{\tiny EMS}}\delta^{jk}-\partial^jU_{\text{\tiny EMS}}\partial^kU_{\text{\tiny EMS}}\Big)\\
\nonumber
&+\frac{1}{c^2}\Big(2\partial^jU_{\text{\tiny EMS}}\partial^kU-\partial^nU_{\text{\tiny EMS}}\partial_nU\delta^{jk}\Big)\bigg]+O(c^{-1}),
\end{align}
\end{subequations}
for the PN expansion of the Landau-Lifshitz pseudotensor components. The last task here is to evaluate $(-g^{(1)})t_{(1)\text{H}}^{\alpha\beta}$. To do so, we substitute Eqs. \eqref{h00_N1}, \eqref{h0j_N1}, and \eqref{hjk_N1} into Eq. \eqref{t_H}. After some simplifications, we deduce that
\begin{subequations}
\begin{align}
\label{t^00_1_H}
&  (-g^{(1)})t_{(1)\text{H}}^{00}=O(1),\\
&  (-g^{(1)})t_{(1)\text{H}}^{0j}=O(c^{-1}),\\
\label{t^jk_1_H}
& (-g^{(1)})t_{(1)\text{H}}^{jk}=\frac{c^4f_0'^2}{\pi G}\bigg[\partial^jU_{\text{\tiny EMS}}\partial^kU_{\text{\tiny EMS}}\\\nonumber
&~~~~~~~~~~~~~~~~~~~~-U_{\text{\tiny EMS}}\partial_n\partial^nU_{\text{\tiny EMS}}\delta^{jk}\bigg]+O(c^{-1}).
\end{align}
\end{subequations}

Gathering together Eqs. \eqref{T00_1}-\eqref{T^jk_1}, and \eqref{T^00_1_EMS}-\eqref{t^jk_1_H}, one can construct the components of the EMSG effective pseudotensor as follows:
\begin{subequations}
\begin{align}
\label{tau_00_EMSG_1}
& c^{-2}\tau^{00}_{(1)\text{eff}}=\rho^*+c^2f_0'{\rho^*}^2+O(c^{-2}),\\
\label{tau_0j_EMSG_1}
& c^{-1}\tau^{0j}_{(1)\text{eff}}=\rho^*v^j+2c^2f_0'{\rho^*}^2v^j+O(c^{-2}),\\
\nonumber
& \tau^{jk}_{(1)\text{eff}}=\rho^*v^jv^k+p\delta^{jk}+\frac{1}{4 \pi G}\bigg[\partial^jU\partial^kU-\frac{1}{2}\partial_nU\partial^nU\delta^{jk}\bigg]\\
\nonumber
&+c^4f_0'\bigg[{\rho^*}^2\delta^{jk}+\frac{1}{c^2}\Big({\rho^*}^2\delta^{jk}\big(2\Pi-v^2-4U\big)+2{\rho^*}^2v^jv^k\\\label{tau_jk_EMSG_1}
&+\frac{1}{\pi G}\big(2\partial^jU_{\text{\tiny EMS}}\partial^kU-\partial_nU_{\text{\tiny EMS}}\partial^nU\delta^{jk}\big)\Big)\\\nonumber
&+f_0'\delta^{jk}\Big(\frac{1}{\pi G}\partial_nU_{\text{\tiny EMS}}\partial^nU_{\text{\tiny EMS}}-4{\rho^*}^2U_{\text{\tiny EMS}}\Big)\bigg]+O(c^{-1}).
\end{align}
\end{subequations}
Here, to simplify the last component, we utilize the Poisson equation $\nabla^2 U_{\text{\tiny EMS}}=-4\pi G {\rho^*}^2$. 
As seen, except for the space-space component, the standard and EMSG parts of $\tau^{00}_{(1)\text{eff}}$ and those of $\tau^{0j}_{(1)\text{eff}}$ have the same order of the magnitude.
Only the term $c^4f_0'{\rho^*}^2$ in $\tau^{jk}_{(1)\text{eff}}$ has an unusual PN order, $O(c^{2})$, compared to the standard terms in this component. In the following computations, we investigate the possible role of this term in the final result.

So far, we have found some source terms which we need to build the gravitational potential tensor in the second iterated step. This tensor in fact has two portions. One part comes from the near-zone solution of the wave equation \eqref{wave_eq} and another one is the wave-zone solution of this equation. The general form of these solutions is given in Appendix. \ref{app_1}.

\subsubsection*{1. Near-zone portion}

Here, we choose the field point within the wave zone and the source point within the near zone. For this case, the general solution form of $ h_{\mathcal{N}}^{\alpha\beta}$ is given in \eqref{hNwave}. We should also recall that according to what was mentioned in Sec. \ref{flux of the energy}, we finally aim to obtain $t_{\text{LL}}^{0k}$, which appears in the energy-balance equation. Therefore, we just need to find the space-space component of $ h_{(2)}^{\alpha\beta}$, cf. Eq. \eqref{t_LL}. 
Rewriting Eq. \eqref{hNwave} for $ h_{(2)\mathcal{N}}^{jk}$, we have
 \begin{align}\label{h2Nwave}
 h_{(2)\mathcal{N}}^{jk}(t,\bm{x})=&\frac{4G}{c^4}\sum_{l=0}^{\infty}\frac{(-1)^{l}}{l!}\\\nonumber
& \times\partial_{j_1j_2\cdots j_l}\bigg[\frac{1}{r}\int_{\mathcal{M}}\tau^{jk}_{\text{eff}}(\tau,\bm{x}')x'^{j_1j_2\cdots j_l}d^3x'\bigg].
 \end{align}

As this step is the very last iterated step of this derivation, we can now apply the gauge condition $\partial_{\beta}h^{\alpha\beta}_{(2)}=0$ or equivalently the conservation equation $\partial_{\beta}\tau^{\alpha\beta}_{\text{(1)eff}}=0$ in order to simplify the general solution \eqref{h2Nwave}.   
In fact, in the final step where the metric truly describes the spacetime with sufficient PN corrections, we enforce the gauge condition/conservation equation to investigate the dynamical behavior of the system \cite{poisson2014gravity}.
Utilizing $\partial_{\beta}\tau^{\alpha\beta}_{\text{(1)eff}}=0$, one can then show that \footnote{The GR versions of the identities \eqref{eq1} and \eqref{eq2} are introduced in chapter 7 of \cite{poisson2014gravity}.}
\begin{subequations}
\begin{align}
\nonumber
&\tau^{jk}_{\text{(1)eff}}=\frac{1}{2}\partial_{00}\big(\tau^{00}_{\text{ (1)eff}}x^jx^k\big)+\frac{1}{2}\partial_q\big(2\tau^{q(j}_{\text{ (1)eff}}x^{k)}\\\label{eq1}
&-\partial_p\tau^{qp}_{\text{(1)eff}}x^jx^k\big),\\
\nonumber
&\tau^{jk}_{\text{(1)eff}}x^p=\frac{1}{2}\partial_0\Big(2\tau^{0(j}_{\text{ (1)eff}}x^{k)}x^p-\tau^{0p}_{\text{(1)eff}}x^jx^k\Big)\\\label{eq2}
&+\frac{1}{2}\partial_n\Big(2\tau^{n(j}_{\text{(1)eff}}x^{k)}x^p-\tau^{pn}_{\text{ (1)eff}}x^jx^k\Big).
\end{align}
\end{subequations}

Now, we return to Eq. \eqref{h2Nwave}. For the case $l=0$, after replacing the integrand in Eq. \eqref{h2Nwave} with Eq. \eqref{eq1}, we arrive at 
\begin{align}
\label{l=0}
\frac{4G}{c^4}&\frac{1}{r}\int_{\mathcal{M}}\tau^{jk}_{\text{ (1)eff}}(\tau,\bm{x}')d^3x'=\\\nonumber
&\frac{2G}{c^4}\frac{1}{r}\partial_{tt}\int_{\mathcal{M}}c^{-2}\tau^{00}_{\text{ (1)eff}}x'^jx'^kd^3x'\\\nonumber
&+\frac{2G}{c^4}\frac{1}{r}\oint_{\partial\mathcal{M}}\Big(2\tau^{q(j}_{\text{ (1)eff}}x'^{k)}-\partial_p\tau^{qp}_{\text{ (1)eff}}x'^jx'^k\Big)dS_q,
\end{align}
where the Gauss divergence theorem has been used.
We first focus our attention on the surface integrals.
As mentioned before, the matter portion of our system is completely restricted to the near zone, and consequently $T^{\alpha\beta}$ and $T^{\alpha\beta}_{\text{\tiny EMS}}$ are zero on  the boundary between the near and wave zones, i.e., the surface $\partial\mathcal{M}$. So, the surface integrals in the above relation reduce to
\begin{align}\label{eq3}
\oint_{\partial\mathcal{M}}\Big(2\tau^{q(j}_{\text{ (1)eff}}[\text{F}]x'^{k)}-\partial_p\tau^{qp}_{\text{(1)eff}}[\text{F}]x'^jx'^k\Big)dS_q,
\end{align}
where $\tau^{jk}_{\text{eff}}[\text{F}]$ representing the field portion of the effective energy-momentum tensor is given by
\begin{align}\label{tau[F]}
\nonumber
& \tau^{jk}_{\text{(1)eff}}[F]=\frac{1}{4 \pi G}\bigg[\partial^jU\partial^kU-\frac{1}{2}\partial_nU\partial^nU\delta^{jk}\bigg]\\
&+\frac{c^4f_0'}{\pi G}\bigg[\frac{1}{c^2}\Big(2\partial^jU_{\text{\tiny EMS}}\partial^kU-\partial_nU_{\text{\tiny EMS}}\partial^nU\delta^{jk}\Big)\\\nonumber
&+f_0'\Big(\partial_nU_{\text{\tiny EMS}}\partial^nU_{\text{\tiny EMS}}-U_{\text{\tiny EMS}}\partial_n\partial^nU_{\text{\tiny EMS}}\Big)\delta^{jk}\bigg]+O(c^{-1}).
\end{align}
In fact, this pseudotensor is entirely constructed from $(-g^{(1)})t_{(1)\text{LL}}^{jk}$ and $(-g^{(1)})t_{(1)\text{H}}^{jk}$ which can exist beyond the near zone, cf.  Eqs. \eqref{t^jk_1_LL} and \eqref{t^jk_1_H}. As seen, the Newtonian gravitational potential $U$ and the EMSG gravitational potential  $U_{\text{\tiny EMS}}$ appear in this source term.
So, to derive the surface integral, let us estimate the order of magnitude of $U$ and $U_{\text{\tiny EMS}}$ on the surface region $\partial\mathcal{M}$. 
In Appendix. \ref{app_3}, by applying the N-body description of the fluid system, we find two expansions \eqref{U_N-body} and \eqref{U_EMSG_N-body} in terms of the distance to the field point for $U$ and $U_{\text{\tiny EMS}}$, respectively. In fact, using these relations, one can approximately evaluate the Newtonian and EMSG gravitational potentials of $N$ separated bodies at a considerable distance from each body. 
Taking advantage of this description and using Eq. \eqref{U_N-body}, we deduce that $U=O(\mathcal{R}^{-1})+O(\mathcal{R}^{-3})+\cdots$ and $\partial_jU=O(\mathcal{R}^{-2})+O(\mathcal{R}^{-4})+\cdots$ on the surface $\partial\mathcal{M}$.
For the EMSG potential, using Eq. \eqref{U_EMSG_N-body}, we also find that $U_{\text{\tiny EMS}}=O(\mathcal{R}^{-1})+O(\mathcal{R}^{-3})+\cdots$ and $\partial_jU_{\text{\tiny EMS}}=O(\mathcal{R}^{-2})+O(\mathcal{R}^{-4})+\cdots$ on the surface $\partial\mathcal{M}$.

Keeping this fact in mind and substituting the first term of Eq. \eqref{tau[F]} into the first part of the surface integral \eqref{eq3}, we have
\begin{align}
\nonumber
\frac{1}{2\pi G}&\oint_{\partial\mathcal{M}}\Big(\partial^qU\partial^{(j}Ux'^{k)}\Big)dS_q =\frac{2}{G}\mathcal{R}^3\langle\partial^qU\partial^{(j}U n^{k)}n_q \rangle\\
&=\frac{2}{G}\mathcal{R}^{-1}\langle n^{(j} n^{k)} \rangle+O(\mathcal{R}^{-m}),
\end{align}
in which $m>1$. Here, we use that $dS_q=\mathcal{R}^2n_q d\Omega$ and $x^k=\mathcal{R}n^k$ on the surface $\partial\mathcal{M}$. We also use this fact that $n^qn_q=1$. In the above relation, the angular bracket denotes an average over the surface  of a sphere, i.e., $\langle\cdots\rangle=(4\pi)^{-1}\int(\cdots)d\Omega$.
As seen, this term is a function of $\mathcal{R}$. 
It is argued that the $\mathcal{R}$-dependent terms in the gravitational potential should be omitted from the main results \cite{poisson2014gravity}. In fact, $\mathcal{R}$-dependent terms in the near-zone portion of $h^{\alpha\beta}$ are finally canceled by those in the wave-zone portion of $h^{\alpha\beta}$. Therefore, we implement this scheme in our calculation and freely discard this term from our results. 
We use the same analysis for the other terms in Eq. \eqref{tau[F]}.
It should be mentioned that here, we use the fact that the angular average of the odd number of $\bm{n}$s is zero.  
Furthermore, as we aim to obtain the T-T part of the gravitational potential $h^{jk}$, we can freely drop terms that are proportional to $\delta^{jk}$.  Finally we find that the surface integrals in Eq. \eqref{l=0} all are a function of $\mathcal{R}$ and consequently have no contribution to the case $l=0$ in the expansion of the gravitational potential $h_{(2)\mathcal{N}}^{jk}(t,\bm{x})$. We leave deriving the volume integral in Eq. \eqref{l=0} here and turn to find the next case $l=1$.

To examine the next term in the expansion of  $h_{(2)\mathcal{N}}^{jk}$,
we set $l=1$ in Eq. \eqref{h2Nwave}. For this term, we have
 \begin{align}\label{eq_l=1}
 \nonumber
&-\frac{4G}{c^4}\partial_{p}\bigg[\frac{1}{r}\int_{\mathcal{M}}\tau^{jk}_{\text{ (1)eff}}(\tau,\bm{x}')x'^{p}d^3x'\bigg]=\\\nonumber
&
-\frac{2G}{c^4}\partial_{p}\bigg[\frac{1}{r}\partial_t\int_{\mathcal{M}}\big(2c^{-1}\tau^{0(j}_{\text{ (1)eff}}x'^{k)}x'^p-c^{-1}\tau^{0p}_{\text{ (1)eff}}x'^jx'^k\big)d^3x'\\
&+\frac{1}{r}\oint_{\partial\mathcal{M}}\Big(2\tau^{n(j}_{\text{ (1)eff}}x'^{k)}x'^p-\tau^{pn}_{\text{(1)eff}}x'^jx'^k\Big)dS_n\bigg].
 \end{align}
Now, we try to obtain the first volume integral in the above relation. As mentioned earlier, our goal is to find $h_{(2)\mathcal{N}}^{jk}(t,\bm{x})$ to the leading PN order. So, the PN expansion of  $c^{-1}\tau^{0j}_{\text{(1)eff}}$ is limited to $O(c^{-2})$. This component is obtained to the required order in Eq. \eqref{tau_0j_EMSG_1}.
We start our calculation in this part by finding the order of magnitude of the first term in the above relation.  In this case, after inserting $\rho^*v^j$ into the volume integral of Eq. \eqref{eq_l=1}, one can easily grasp that this volume integral is proportional to $I^{jkp}=\int \big(2\rho^*v^{(j}x^{k)}x^p-\rho^*v^px^jx^k\big)d^3x$. Regarding this relation, we deduce that $I^{jkp}$ is of the order  $O(m_cv_cr_c^2)$. 
Here, the quantities $m_c$ and $r_c$ represent the characteristic mass and length scale of the system, respectively. Moreover, $v_c$ is the characteristic velocity of the system. 
So, we deduce that the time derivation of this term is of the order $\dot{I}^{jkp}=O(\frac{m_cv_cr_c^2}{t_c})$ in which $t_c$ is the characteristic time scale during which the source changes. In the other words, $t_c=r_c/v_c$.
We should recall that $I^{jkp}$ is a function of the retarded time $\tau$.
Therefore, the first term in Eq. \eqref{eq_l=1} becomes
\begin{align}
&-\frac{2G}{c^4}\partial_p(\frac{1}{r}\dot{I}^{jkp})=\frac{2G}{c^4}\Big(\frac{1}{cr}\ddot{I}^{jkp}n_p+\frac{1}{r^2}\dot{I}^{jkp}n_p\Big)\\
&=O\Big(\frac{G}{c^5}\frac{m_cv_cr_c^2}{rt_c^2}\big(1+\frac{\lambda_c}{r}\big)\Big)\simeq O\Big(\frac{G}{c^5}\frac{m_cv_cr_c^2}{rt_c^2}\Big).
\end{align}
To simplify the above relation, we use $\partial \tau/\partial x^p=-c^{-1}n_p$ and also consider that $\lambda_c\ll r$ where $\lambda_c=c\, t_c$ is the characteristic wavelength of the radiation. This condition reveals that the field point is situated in the wave zone. In fact, as we aim to study the gravitational potential in the wave zone, the assumption $\lambda_c\ll r$ is completely suitable here.  
Now, we compare this term with the volume integral in the case $l=0$.  By substituting the leading term of Eq. \eqref{tau_00_EMSG_1} within the volume integral of Eq. \eqref{l=0}, we evaluate the order of magnitude of this term as 
\begin{align}\label{tau00leading}
\frac{4G}{c^4}\frac{1}{r}\partial_{tt}\int_{\mathcal{M}}{\rho^*}'x'^jx'^kd^3x'=O\Big(\frac{G}{c^4}\frac{m_cr_c^2}{rt_c^2}\Big).
\end{align}
From these estimations, one can easily grasp that the volume integral in the $l=1$ case compared to the leading term in the $l=0$ case is of the order $O(v_c/c)$.
By applying a similar analysis, we find that the second term in Eq. \eqref{tau_0j_EMSG_1} compared to Eq. \eqref{tau00leading} is also of the order $O(v_c/c)$.
Therefore, this term in total will add a $0.5$\tiny PN \normalsize correction to the PN expansion of the gravitational potential. On the other hands, if the leading term of the $l=0$ case is considered as the $1$\tiny PN \normalsize correction, the biggest contribution of the $l=1$ case to Eq. \eqref{h2Nwave} will be of the $1.5$\tiny PN \normalsize order. 

As the final part of our calculation in this subsection, it is important to obtain the surface integral in the $l=1$ case, cf. Eq. \eqref{eq_l=1}. To do so, we again restrict the source of this surface integral to the field portion of the effective energy-momentum tensor introduced in Eq. \eqref{tau[F]}. So, we have
\begin{align}
\oint_{\partial\mathcal{M}}\Big(2\tau^{n(j}_{\text{(1)eff}}[F]x'^{k)}x'^p-\tau^{pn}_{\text{ (1)eff}}[F]x'^jx'^k\Big)dS_n,
\end{align}
for this surface integral. By applying a similar analysis mentioned above, one can show that this integral is simplified to the angular average of the odd number of $\bm{n}$s times the $\mathcal{R}$-dependent terms. So, this surface integral eventually vanishes and has no contribution to $h_{(2)\mathcal{N}}^{jk}$.
It should be emphasized that the unusual high-order term appearing in the space-space component of $\tau_{(1)\text{eff}}^{\alpha\beta}$ just contributes to the surface integrals. Completely depending to the matter part of the system, this term then has no role up to the leading order. 
To sum up, we deduce that each $l$-pole term in Eq. \eqref{h2Nwave} will be the $(l/2)$\tiny PN \normalsize correction in the PN expansion of the near-zone gravitational potential.

Considering the above discussion, we finally deduce that the PN expansion of the near-zone gravitational potential \eqref{h2Nwave} is simplified as
\begin{align}\label{h_2_final}
h_{(2)\mathcal{N}}^{jk}=\frac{2G}{c^4}\frac{\ddot{\mathcal{I}}^{jk}_{(2)}(\tau)}{r}+O(c^{-5}),
\end{align}
where
\begin{align}\label{I^jk}
\mathcal{I}^{jk}_{(2)}=\int_{\mathcal{M}}c^{-2}\tau^{00}_{\text{ (1)eff}}(\tau,\bm{x}')x'^jx'^kd^3x'.
\end{align}
To find the leading PN order of Eq. \eqref{h_2_final} and consequently the leading term in the energy-balance equation, we should only insert the order $c^{2}$ of $\tau^{00}_{\text{(1)eff}}$
into the definition \eqref{I^jk}.
This order is derived in the previous part. 
As seen, in comparison with the GR system, for two cases $f_0'>0$ or $f_0'<0$,  the gravitational potential can be stronger or weaker in EMSG, respectively.  
So far, we have derived the near-zone portion of $h_{(2)}^{jk}$ described by Eqs. \eqref{h_2_final}. To complete our calculation, we obtain the wave-zone portion of the gravitational potential in the rest of this section.

\subsubsection*{2. Wave-zone portion}

In order to find the wave-zone portion of the total gravitational potential, $h^{jk}_{(2)\mathcal{W}}$,  enough information about the source terms, cf. Eq. \eqref{tau}, is required. It means we should obtain the effective energy-momentum pseudotensor $\tau^{\alpha\beta}_{\text{(1)eff}}$ in the wave zone. We recall that in the previous part, we obtained this pseudotensor in the near zone. To construct $\tau^{\alpha\beta}_{\text{(1)eff}}$ existing in the wave zone, we first calculate its foundations, i.e., $h_{(1)}^{\alpha\beta}$ in which the field point is situated in the wave zone.  
By using relations mentioned in Appendix. \ref{app_1}, we find $h_{(1)}^{\alpha\beta}$ where $\vert \bm{x} \vert \geq\mathcal{R}$ as follows. After substituting Eqs. \eqref{T(0)} and \eqref{T00(0)EMS} within Eq. \eqref{hNwave}, we have
\begin{align}\label{h00_1N}
h_{(1)\mathcal{N}}^{00}=\frac{4G}{c^2}\frac{1}{r}\Big(M_0(\tau)+c^2f_0'\mathfrak{M}(\tau)\Big)+O(c^{-3}),
\end{align}
for the time-time component of the gravitational potential in the first iteration of the wave equation. Here,  
\begin{align}
M_0=\int_{\mathcal{M}}{\rho^*}d^3x',
\end{align}
in principle, represents the total matter inside the near zone.
As before, we consider that $f_0'$ is a constant. In the above relation, one can show that $O(c^{-3})$ comes from $l=1$ terms in the expansion \eqref{hNwave}.
In a similar way, for the time-space component of the potential, we arrive at 
\begin{align}\label{h0j_1N}
h_{(1)\mathcal{N}}^{0j}=\frac{4G}{c^3}\frac{1}{r}\Big(P_0^j(\tau)+2c^2f_0'\,\mathfrak{P}^j(\tau)\Big)+O(c^{-4}),
\end{align}
where
\begin{align}
P_0^j=\int_{\mathcal{M}}{\rho^*}v^jd^3x',
\end{align}
and 
\begin{align}\label{new}
\mathfrak{P}^j=\int_{\mathcal{M}}{{\rho^*}'}^2v'^jd^3x'.
\end{align}
In this component, $l=1$ terms contribute at $O(c^{-4})$ and its $O(c^{-3})$ term is purely built of the $l=0$ pole. 
Finally, for the space-space component, we obtain 
\begin{align}\label{hjk_1N}
h_{(1)\mathcal{N}}^{jk}=4G\frac{f_0' }{r} \mathfrak{M}\, \delta^{jk}+O(c^{-3}).
\end{align}
Notice $M_0$, $P_0^j$, $\mathfrak{M}$, and $\mathfrak{P}^j$ all are functions of the retarded time $\tau$ here.
In this component, the  $c^{-3}$ order is also constructed from $l=1$ terms of the expansion \eqref{hNwave}. In general, one can show that each $l$-pole term in this expansion adds an $l/2$\tiny PN \normalsize contribution to $h^{\alpha\beta}$. 
Until now, we derive the near-zone portion of the gravitational potential, i.e., $h_{(1)\mathcal{N}}^{\alpha\beta}$. To complete the gravitational potential, we should also find $h_{(1)\mathcal{W}}^{\alpha\beta}$ where both the field and source points are in the wave zone. On the other hand, regarding Eqs. \eqref{T(0)}-\eqref{T(0)EMSG}, one can easily conclude that $h_{(1)\mathcal{W}}^{\alpha\beta}$ vanishes and then $h_{(1)}^{\alpha\beta}=h_{(1)\mathcal{N}}^{\alpha\beta}$. As seen in this part of the derivation, in the EMSG theory, the gravitational potential can exist in the wave zone even at the first iteration step. At the first glance, it shows that one can study the energy-balance equation with the information derived in the first iteration. However, considering the coefficient $\delta^{jk}$ in Eq. \eqref{hjk_1N}, this potential in fact has no T-T portion to construct the Landau-Lifshitz pseudotensor at this order, cf. Eqs. \eqref{eq01} and \eqref{t_LL}. From this point, we continue our computation until the second iteration. 

Now, applying the components of $h_{(1)\mathcal{N}}^{\alpha\beta}$, we can find $\tau_{\text{(1)eff}}^{\alpha\beta}$ in the wave zone. We first turn to find the Landau-Lifshitz pseudotensor $(-g^{(1)})t_{(1)\text{LL}}^{\alpha\beta}$. 
To do so, we insert Eqs. \eqref{h00_1N}, \eqref{h0j_1N}, and \eqref{hjk_1N} into definition \eqref{tLL_harmonic}. After some simplifications, we arrive at
\begin{subequations}
\begin{align}
\nonumber
&(-g^{(1)})t_{(1)\text{LL}}^{00}=-\frac{G}{\pi r^4}\Big(\frac{7}{8}M_0^2+c^4f_0'\mathfrak{M}\big(f_0'\mathfrak{M}\\\label{eq4}
&~~~~~~~~~~~~~~~~~~~~~+\frac{1}{c^2}M_0\big)\Big)+O(c^{-1}),\\
&
(-g^{(1)})t_{(1)\text{LL}}^{0j}=O(c^{-1}),\\
&
(-g^{(1)})t_{(1)\text{LL}}^{jk}=\frac{G}{\pi r^4}\bigg[\Big(\frac{1}{4}M^2_0-c^4f_0'\mathfrak{M}\big(f_0'\mathfrak{M}\\\nonumber
&-\frac{2}{c^2}M_0\big)\Big)\Big(n^jn^k-\frac{1}{2}\delta^{jk}\Big)+\frac{1}{2}c^4{f_0'}^2\mathfrak{M}^2\delta^{jk}\bigg]+O(c^{-1}).
\end{align}
\end{subequations}
We will see that the PN order of the space-space component is indeed sufficient to calculate the 
$h_{(2)\mathcal{W}}^{jk}$ to the leading order. 
Here, we use this fact that $\partial_j \tau=-1/c\, n^j$. We also use that $\partial_j F(\tau)=-1/c\, n^j \partial_{\tau} F$ where $F$ is an arbitrary function of the retarded time.
Moreover, due to Eq. \eqref{rho}, one can conclude that $dM_0/dt=0$.  
To find the harmonic portion, we substitute the components of $h_{(1)\mathcal{N}}^{\alpha\beta}$ within the definition \eqref{t_H}. So, we have  
\begin{subequations}
\begin{align}
\nonumber
&(-g^{(1)})t_{(1)\text{H}}^{00}=-\frac{2 G }{\pi r^4}c^4 f_0' \mathfrak{M}\big(f_0'\mathfrak{M}+\frac{1}{c^2}M_0\big)\\
&~~~~~~~~~~~~~~~~~~~~+O(c^{-1}),\\
&(-g^{(1)})t_{(1)\text{H}}^{0j}=O(c^{-1}),\\
\nonumber
&(-g^{(1)})t_{(1)\text{H}}^{jk}=\frac{G}{\pi r^4}c^4 {f_0'}^2 \mathfrak{M}^2\big(n^jn^k-2\delta^{jk}\big)\\\label{eq5}
&~~~~~~~~~~~~~~~~~~~~+O(c^{-1}).
\end{align}
\end{subequations}
Here, we use that $\partial_j n_k=\frac{1}{r}\big(\delta_{jk}-n_jn_k\big)$. 
On the other hand, since the matter part of the system cannot exist in the wave zone, the entire energy-momentum pseudotensor $\tau_{\text{(1)eff}}^{\alpha\beta}$ in the wave zone is constructed from $(-g^{(1)})t_{(1)\text{LL}}^{\alpha\beta}$ and $(-g^{(1)})t_{(1)\text{H}}^{\alpha\beta}$. Regarding this point and gathering together Eqs. \eqref{eq4}-\eqref{eq5}, we finally arrive at
\begin{subequations}
\begin{align}
\nonumber
&\tau_{\text{(1)eff}}^{00}=-\frac{G}{\pi r^4}\Big(\frac{7}{8}M_0^2+3 c^4f_0'\mathfrak{M}\big(f_0'\mathfrak{M}+\frac{1}{c^2}M_0\big)\Big)\\
&~~~~~~~~~~~+O(c^{-1}),\\
& \tau_{\text{(1)eff}}^{0j}=O(c^{-1}),\\
\nonumber
& \tau_{\text{(1)eff}}^{jk}=\frac{G}{\pi r^4}\Big[\big(\frac{1}{4}M^2_0+2c^2f_0'\mathfrak{M}M_0\big)\big(n^jn^k-\frac{1}{2}\delta^{jk}\big)\\\label{eq6}
&~~~~~~~~~~~-c^4{f_0'}^2\mathfrak{M}^2\delta^{jk}\Big]+O(c^{-1}).
\end{align}
\end{subequations}
It should be mentioned that, here, we only need to obtain the space-space component of $\tau_{\text{(1)eff}}^{\alpha\beta}$. However, for the sake of completeness, we have found the other components as well.

Before deriving $h_{(2)\mathcal{W}}^{jk}$, let us simplify the source term of this potential, i.e, $\tau_{\text{(1)eff}}^{jk}$ so that the symmetric trace-free tensors appear in it. To do so, we rewrite Eq. \eqref{eq6} as follows:
\begin{align}
\nonumber
&\tau_{\text{(1)eff}}^{jk}=\frac{G}{\pi r^4}\Big[\big(\frac{1}{4}M^2_0+2c^2f_0'\mathfrak{M}M_0\big)\big(n^{<jk>}-\frac{1}{6}\delta^{jk}\big)\\
&-c^4{f_0'}^2\mathfrak{M}^2\delta^{jk}\Big]+O(c^{-1}),
\end{align}
in which $n^{<jk>}=n^jn^k-1/3\delta^{jk}$ is a symmetric trace-free tensor. Now, by comparing this relation with Eq. \eqref{eq7}, we can simply deduce that in this case, the source of the gravitational potential $h_{(2)\mathcal{W}}^{jk}$ is a function of $r^{-n}$ with $n=4$ and
\begin{subequations}
\begin{align}
\label{eq9}
& f_{l=0}^{jk}=-\frac{G}{6}\delta^{jk}\Big(M_0^2+8c^4f_0'\mathfrak{M}\big(3 f_0'\,\mathfrak{M}+\frac{1}{c^2} M_0\big)\Big),\\
\label{eq10}
& f_{l=2}^{jk}=G M_0\big(M_0+8c^2f_0' \mathfrak{M}\big).
\end{align}
\end{subequations}
Up to this point, we have achieved adequate materials to calculate the integral \eqref{eq8}. Considering $n=4$ and applying Eqs. \eqref{eq9}-\eqref{eq10} for $l=0$ and $l=2$ respectively, after some simplifications, we eventually arrive at
\begin{align}\label{h_2_W}
\nonumber
h_{(2)\mathcal{W}}^{jk}=&\frac{G^2}{c^4 r^2}\bigg\lbrace\Big(M_0^2+8c^2f_0' \mathfrak{M}M_0\Big)n^jn^k\\
&+8c^4 {f_0'}^2\mathfrak{M}^2\delta^{jk}\bigg\rbrace,
\end{align}
for the space-space component of the gravitational potential in the wave zone. Note that as already mentioned, we drop the $\mathcal{R}$-dependent terms.
It is obvious from this relation that the wave-zone contribution of the gravitational potential is proportional to the inverse squared $r$.
However, because we aim to find the flux of the energy in the far-away wave zone, in comparison to $h_{(2)\mathcal{N}}^{jk}$ which falls as $r^{-1}$, we may discard this contribution to $h^{jk}$. Furthermore, this part is 0.5\tiny PN \normalsize order higher than the leading term in Eq. \eqref{h_2_final}.
To show this fact, regarding Eqs. \eqref{h_2_final} and \eqref{h_2_W}, we deduce that $h_{(2)\mathcal{W}}^{jk}/h_{(2)\mathcal{N}}^{jk}= O(G M t^2_{\text{c}}/r r^2_\text{c})$.
Considering the Newtonian acceleration, one can show that $G M$ is of the $r^3_{\text{c}}/t^2_{\text{c}}$ order. By applying this order for $G M$ and also setting $r\sim \lambda_{\text{c}}=c\,t_{\text{c}}$, one can conclude that $h_{(2)\mathcal{W}}^{jk}/h_{(2)\mathcal{N}}^{jk}=O(v_{\text{c}}/c)$. 
Therefore, this portion of the potential adds a 0.5\tiny PN \normalsize correction to $h_{(2)}^{jk}$, i.e., $h_{(2)\mathcal{W}}^{jk}$ is actually of the 1.5 \tiny PN \normalsize order. To sum up, with regard to both former and latter facts, we can ignore this part in our calculation. So, we have $h_{(2)}^{jk}=h_{(2)\mathcal{N}}^{jk}$.

\section{Gravitational energy flux in EMSG}\label{Gravitational energy flux in EMSG}

It is comprehensively shown that the leading order of the gravitational potential in the EMSG theory is described by Eq. \eqref{h_2_final}  and the only difference with the GR case comes from the definition of quadrupole-moment tensor $\mathcal{I}^{jk}$ in this theory. Rewriting Eq. \eqref{I^jk} up to the leading order, we have
\begin{align}\label{I^jk_leading}
\mathcal{I}^{jk}=\int_{\mathcal{M}}\big(\rho+c^2f_0'\rho^2\big)x^jx^kd^3x.
\end{align}
As seen in addition to the usual term $\rho$, a new quadratic term $\rho^2$ plays a role in this order. Since we focus on the leading order, hereafter, we drop ``$*$" and exhibit the density with $\rho$. Therefore, one can straightforwardly deduce that the EMSG version of the instantaneous gravitational energy flux is given by
\begin{align}
\label{p}
\mathcal{P}=\frac{G}{5c^5}\dddot{\mathcal{I}}^{<jk>}\dddot{\mathcal{I}}_{<jk>},
\end{align}
in which $\mathcal{I}^{<jk>}=\mathcal{I}^{jk}-1/3\,\delta^{jk}\,\mathcal{I}^{qq}$ is the symmetric-tracefree sector of $\mathcal{I}^{jk}$. It should be mentioned that to derive $\mathcal{P}$, the angular average of the radial vectors $\boldsymbol{n}$s is used here. 
The difference with GR is that instead of the Newtonian mass quadrupole moment $I^{jk}$, we insert $\mathcal{I}^{jk}$.  It is obvious that, like in GR (cf. \cite{will2018theory}), the monopole and dipole terms vanish here and in this PN order, the quadrupole term contributes to the flux of energy. 

In order to clarify the role of the EMSG correction in this relation practically, we investigate $\mathcal{P}$ for a binary system containing two compact bodies with mass $m_A$ where the index ``$A$ " stands for the body $A$ and $A=1,2$. As usual, we choose the barycenter coordinate system to study the binary system. In this coordinate system, the position of each body $\boldsymbol{r}_A(t)$ is given with respect to the system's barycenter. We also have
\begin{subequations}
\begin{align}
\label{r_1andr_2}
&\boldsymbol{r}_1=\frac{m_2}{m}\boldsymbol{r},~~~~~~~~\boldsymbol{r}_2=-\frac{m_1}{m}\boldsymbol{r},\\
&\boldsymbol{v}_1=\frac{m_2}{m}\boldsymbol{v},~~~~~~~~\boldsymbol{v}_2=-\frac{m_1}{m}\boldsymbol{v},
\end{align}
\end{subequations}
where $m=m_1+m_2$ is total mass of the system. Here, $\boldsymbol{r}=\boldsymbol{r}_1-\boldsymbol{r}_2$ is the separation vector of the bodies and $\boldsymbol{v}=d\boldsymbol{r}/dt$ is their relative velocity. 
It should be noted, because of the presence of the quadratic term $\rho^2$ in the integral $\mathcal{I}^{jk}$, we cannot simply utilize the point-mass description and take the advantage of the delta function $\delta(\boldsymbol{x})$ to solve this integral. In this case, the integral \eqref{I^jk_leading} diverges. Therefore, to carry out our calculation, we consider compact bodies that are well separated. In the framework of this description, we have $\rho=\sum_A\rho_A$ in which $\rho_A$ is the density of the body $A$ and vanishes outside the volume $V_A$ occupied by this body.  We also introduce the vector $\bar{\boldsymbol{x}}=\boldsymbol{x}-\boldsymbol{r}_A$ that shows the fluid element position with respect to the center of mass of the body $A$. This description is also mentioned in Appendix. \ref{app_3}. Applying these definitions, for the second term of Eq. \eqref{I^jk_leading}, we have 
\begin{align}
\nonumber
\int_{\mathcal{M}}\rho^2x^jx^kd^3x&\simeq\int_{\mathcal{M}_1}\rho^2_1\big(\bar{x}^j+r_1^j\big)\big(\bar{x}^k+r_1^k\big)d^3\bar{x}\\
&+\int_{\mathcal{M}_2}\rho^2_2\big(\bar{x}^j+r_2^j\big)\big(\bar{x}^k+r_2^k\big)d^3\bar{x}.
\end{align}
In the above relation, the domain of the first integral is a sphere centered at the binary system's center of mass with a radius $\vert \boldsymbol{x}\vert <\mathcal{R}$.  
One can approximately simplify this integral in terms of the new domain $\mathcal{M}_A$ which this time is a sphere with a radius $\vert \bar{\boldsymbol{x}}\vert <\mathcal{R}$ and its center corresponds to the body $A$'s center of mass. As the integrands are completely related to the matter sector, these domains are eventually restricted to the volumes occupied by these bodies. 
Therefore, we arrive at 
\begin{align}
\nonumber
\int_{\mathcal{M}}\rho^2x^jx^kd^3x\simeq  \mathfrak{M}_{1}r_1^jr_1^k&+ \mathfrak{M}_{2}r_2^jr_2^k\\
&+\mathfrak{I}^{jk}[\text{cm}_1]+\mathfrak{I}^{jk}[\text{cm}_2],
\end{align}
where $\mathfrak{M}_A$ and $\mathfrak{I}^{jk}[\text{cm}_A]$ are given by Eqs. \eqref{M_A} and \eqref{I_A^jk}. 
Here, ``$\text{cm}_A$" shows that the mentioned quantity is obtained with respect to the center of mass of the body $A$.
To simplify this relation, we consider the symmetry $\rho(t,\boldsymbol{r}_A-\bar{\boldsymbol{x}})=\rho(t,\boldsymbol{r}_A+\bar{\boldsymbol{x}})$. 
Following a similar computation for the first term in Eq. \eqref{I^jk_leading}, we have
\begin{align}
\nonumber
\int_{\mathcal{M}} \rho\, x^jx^kd^3x\simeq m_1r_1^jr_1^k&+m_2 r_2^jr_2^k\\
& +I^{jk}[\text{cm}_1]+I^{jk}[\text{cm}_2],
\end{align}
in which $I^{jk}[\text{cm}_A]=\int_A\rho_A \bar{x}^j\bar{x}^kd^3\bar{x}$ is the quadrupole-moment tensor relative to the center of mass of the body $A$. Gathering together the above results and applying Eq. \eqref{r_1andr_2}, we obtain $\mathcal{I}^{jk}$ for the binary system as
\begin{align}\label{I^jk_binary}
\mathcal{I}^{jk}&=\Big(\eta m+c^2f_0'\frac{1}{m^2}\big(\mathfrak{M}_1m_2^2+\mathfrak{M}_2m_1^2\big)\Big)r^jr^k\\\nonumber
&+I^{jk}[\text{cm}_1]+I^{jk}[\text{cm}_2]+c^2f_0'\big(\mathfrak{I}^{jk}[\text{cm}_1]+\mathfrak{I}^{jk}[\text{cm}_2]\big),
\end{align} 
where $\eta=m_1m_2/m^2$.  

The next step toward finding $\mathcal{P}$ is to compute the time derivative of this quadrupole-moment tensor. To do so, we consider that the temporal variations of the internal tensors $I^{jk}[\text{cm}_A]$ and $\mathfrak{I}^{jk}[\text{cm}_A]$ are insignificant compared to those related to the orbital motion. Consequently, we set $\dot{I}^{jk}[\text{cm}_A]=0=\dot{\mathfrak{I}}^{jk}[\text{cm}_A]$. For the sake of simplification, we also assume that the density inside each body is uniform so that $\boldsymbol{\nabla}\rho_A=0$. Under this condition, given the result of Appendix. \ref{app_2}, we find that $d \mathfrak{M}_A/dt=0$. Moreover, one can simplify the second term in Eq. \eqref{I^jk_binary} as $\big(\mathfrak{M}_1m_2^2+\mathfrak{M}_2m_1^2\big)/m^2=\eta\big(\rho_1m_2+\rho_2m_1\big)$.
Therefore, to complete our derivation, we only need to compute $d^3(r^jr^k)/dt^3$. Since, we aim to obtain $\mathcal{P}$ to the leading order, we can use the Keplerian equations to describe the motion of the binary system. We consider that the orbit is in the $(x,y)$ plane. Consequently, we have
\begin{align}
\nonumber
\frac{d^3}{dt^3}\big(r^jr^k\big)&=-2\frac{(G m)^{\frac{3}{2}}}{l^{\frac{5}{2}}}\big(1+e\cos\varphi\big)^2\Big[e\sin\varphi\, n^jn^k\\
&+2\big(1+e\cos\varphi\big)\big(n^j\lambda^k+\lambda^jn^k\big)\Big],
\end{align}
in which $l$
and $e$ are the semilatus rectum and eccentricity of the orbit, respectively. Here, $\boldsymbol{n}$ and $\boldsymbol{\lambda}$ are the orbital plane's unit vectors of the binary system that are given by $\boldsymbol{n}=\big[\cos\varphi,\sin\varphi,0\big]$ and $\boldsymbol{\lambda}=\big[-\sin\varphi,\cos\varphi,0\big]$ where $\varphi$ is the true anomaly. Inserting these results within Eq. \eqref{p}, after some manipulations, we finally arrive at
\begin{align}
\label{p_binary}
&\mathcal{P}=\frac{32}{5}\frac{\eta^2}{G}\Big(\frac{G\, m}{c\, l}\Big)^5\Big(1+2\alpha\Big)\\\nonumber
&\times\big(1+e\cos\varphi\big)^4\Big[1+2e\cos\varphi+\frac{1}{12}e^2\big(1+11\cos^2\varphi\big)\Big],
\end{align}
in which 
\begin{align}\label{EMSG_parameter}
\alpha= \frac{c^2f_0'}{m}\big(\rho_1m_2+\rho_2m_1\big).
\end{align}
Notice that given the EMSG correction $\alpha$ is a small parameter, 
$|\alpha|\ll 1$, we have expanded our results in powers of $\alpha$, kept only linear terms in $\alpha$, and then obtained Eq. \eqref{p_binary}.
This relation describes the flux of energy in the EMSG theory. Obviously, dropping the EMSG correction $\alpha$, the relation \eqref{p_binary} reduces to the GR version.
It should be emphasized that when the density inside each body is not uniform, this relation would dramatically change. In fact, in a more realistic case where $\rho$ is not constant, $d \mathfrak{M}_A/dt$ should be considered in calculating the EMSG energy flux.

\begin{figure}
\begin{center}
\centering
\includegraphics[scale=0.83]{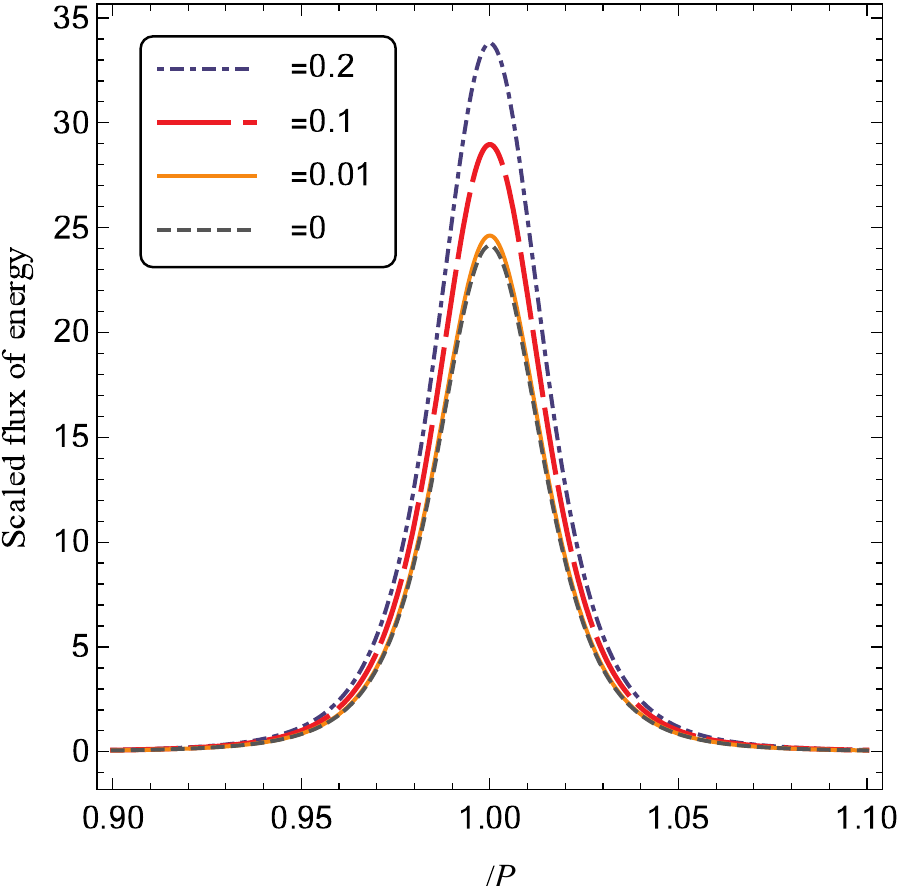}
\caption{The scaled flux of energy $\mathcal{P}(5/32)(G/\eta^2)(c\,l/G\,m)^5$ in terms of the scaled retarded time $\tau/P$ where $P$ is the orbital period. Here, we assume that $e=0.7$. In this figure, $\alpha$ shows the EMSG correction. We see that near the pericenter, even for a rather small value of $\alpha$, the flux of energy changes considerably comparing with the GR case with $\alpha=0$.}\label{fig1}
\end{center}
\end{figure}

In order to visualize the effect of the EMSG correction on $\mathcal{P}$, we display Eq. \eqref{p_binary} in Fig. \ref{fig1}. 
To do so, we choose a specific case $e=0.7$ and then numerically solve $\dot{\varphi}=\big(Gm/l^3\big)^{1/2}(1+e\cos\varphi)^2$ in terms of the retarded time. 
Here, we also assume that the density of the two bodies in the binary system is of the same order and it is constant.
We set $\rho_1=\rho_2=\rho$. Under this assumption, the EMSG correction \eqref{EMSG_parameter} reduces to $\alpha=c^2f_0'\rho$. As seen from this figure, even for the small value of $\alpha$, the flux of energy near the pericenter of the orbit dramatically changes 
comparing with the GR one. Therefore, one may conclude that studying this relation could strongly restrict the free parameter of this theory $f_0'$. 
In the following section, by studying the post-Keplerian parameter, the time derivative of the orbital period $\dot{P}$, \footnote{Notice that hereafter, the symbol $P$ stands for the orbital period and \textit{almost} the same symbol $\mathcal{P}$ represents the flux of energy.} of several binary pulsars, we attempt to find a bound on $f_0'$.

As a final point in this section, let us return to the energy flux derived in Eq. \eqref{p_binary}. As shown, in the EMSG theory, this relation depends on the densities of each component of the binary system in addition to their masses. So, for dense systems like double-neutron star binaries, the EMSG correction will be considerable. On the other hand, it means that the gravitational radiation in this theory may implicitly be affected by the internal structure of each component.
Now the question is whether it can be a consequence of the violation of the Strong Equivalence Principle (SEP) in this theory. To demonstrate the latter state, we consider the near-zone metric obtained in Sec. \ref{first iteration}, cf. Eqs. \eqref{g00_1}-\eqref{gij_1}. According to the relations represented in Appendix. \ref{app_3}, one can show that in the framework of the N-body description, for instance, the time-time component of the spacetime metric outside a single sphere with mass $m$ and constant density $\rho$ is given by  $g_{00}=-1+2G\,m/(rc^2)\big(1+4c^2f_0'\rho\big)$.  It illustrates that two compact stars with the same rest mass $m$ but different densities would induce different spacetimes. In other words, not only their mass but also their density could affect the trajectory of a test particle \cite{Nazari2021bending}.
This reveals that the interior of the bodies could play a role here. Therefore, we conclude
that SEP is indeed violated in this theory.
This is expected in the sense that the standard conservation equation for $T_{\mu\nu}$ does not hold in EMSG. This means that particles do not necessarily move on geodesics. More specifically, the internal properties of the massive bodies appear at the right-hand side of the conservation equation. 
It should be mentioned that to examine the bodies' internal structure on the near-zone spacetime completely, one should at least find the metric up to the second iteration where the required PN order is faithfully constructed.

\section{Application to the binary pulsars}\label{Application to the binary pulsars}

       \begin{table*}[!]
       \caption{The binary pulsars and their characteristics}\label{table1}
       \begin{center}
       \footnotesize
      \renewcommand{\arraystretch}{1.5}
         \begin{tabular}{lcccccccccccccccccccccccc}
         \hline        
         System & && $P $ && & $e$ & && $m_{\text{p}}$ & && $m_{\text{c}}$ & && $\dot{P}_{\text{obs}}$& && $\pm \delta$ && & $\dot{P}_{\text{GR}}$ &&& References\\
           && & $(\text{d})$  && & & && $(M_{\odot})$ & && $ (M_{\odot})$ & && $ (\times10^{-12})$ &&& $(\times10^{-12})$ && & $(\times10^{-12})$ &&& \\
         \hline
         \hline
         J2129+1210C && & $0.335\,282\,049$ & && $0.681\,395$ & && 1.358 & && 1.354 & && $-3.96$ & && 0.05 &&& $-3.94$ &&& \cite{jacoby2006measurement}\\
         J1915+1606 &&& $0.322\,997\,449$ &&& $0.617\,1334$ &&& 1.4398 &&& 1.3886 &&& $-2.423$ &&& 0.001 &&& $-2.403$ &&& \cite{hulse1975discovery,weisberg2010timing} \\
         J0737-3039A &&& $0.102\,251\,562$ &&& $0.087\,7775$ &&& 1.3381 &&& 1.2489 &&& $-1.252$ &&& 0.017 &&& $-1.248$ &&& \cite{kramer2006tests} \\
         J1537+1155 &&& $0.420\,737\,299$ &&& $0.273\,6767$ &&& 1.3332 &&& 1.3452 &&& $-0.138$ &&& 0.0001 &&& $-0.192$ &&& \cite{stairs2002studies}
         \\
         J1141-6545 &&& $0.197\,650\,959$ &&& $0.171\,884$ &&& 1.27 &&& 1.02 &&& $-0.403$ &&& 0.025 &&& $-0.387$ &&& \cite{bhat2008gravitational}\\
         J1738+0333 &&& $0.354\,790\,7399$ &&& $3.4\times10^{-7}$ &&& 1.46 &&& 0.181 &&& $-0.017$ &&& 0.0031 &&& $-0.0277$ &&& \cite{freire2012relativistic}\\
        \hline
         \end{tabular}
       \end{center} 
       \normalsize
       \end{table*}

As an application of our derivation, we investigate the EMSG effect on the time average of the power radiated due to GWs of the binary system. Then, applying this relation, we study the EMSG version of the post-Keplerian parameter $\dot{P}$. As the orbital period variation is the best-observed parameter, we choose this post-Keplerian parameter to find a reasonable restriction on the free parameter of the EMSG theory. Of course, to examine the EMSG relativistic effects comprehensively, the other post-Keplerian parameters should also be studied. Moreover, studying all post-Keplerian parameters, one can check the self-consistency of the theory.
e.g., in the case of the tensor-scalar theories of gravity, see \cite{bhat2008gravitational,freire2012relativistic,damour1996tensor}. We next find $\dot{P}_{\text{\tiny EMSG}}$ of several known binary pulsars; and by comparing our result with their observed orbital period variation $\dot{P}_{\text{obs}}$, we try to set a bound on $f_0'$. 

Regarding Eq. \eqref{p_binary}, we obtain the orbital average of the flux of energy, $\big<\mathcal{P}\big>$, and then according to the energy-balance equation \eqref{energy-balance_equation}, we arrive at  
\begin{align}\label{dE_binary_average}
\nonumber
&\Big<\frac{dE}{dt}\Big>=-\frac{32}{5}\frac{\eta^2}{G}\Big(\frac{G\, m}{c\, l}\Big)^5\Big(1+2\alpha\Big)\\
&\times\big(1-e^2)^{\frac{3}{2}}\Big[1+\frac{73}{24}e^2+\frac{37}{96}e^4\Big],
\end{align}
for the time average of the radiated power. Inserting the orbital energy 
and the Kepler third law into the above relation yields
\begin{align}\label{Tdot}
\nonumber
&\dot{P}_{\text{\tiny EMSG}}=-\frac{192\pi}{5}\Big(\frac{G M}{c^3}\frac{2\pi}{P}\Big)^{\frac{5}{3}}\Big(1+2\alpha\Big)\\
&\times\big(1-e^2)^{-\frac{7}{2}}\Big[1+\frac{73}{24}e^2+\frac{37}{96}e^4\Big],
\end{align}
for the first time derivative of the orbital period $P$. Here, $M=\eta^{3/5}m$ is the chirp mass.  As seen, the quadrupolar gravitational radiation in the EMSG theory is obviously dictated in this post-Keplerian parameter. Therefore, this relativistic parameter can be a sound test bed for this theory.

Given the sign of $f_0'$, $\dot{P}_{\text{\tiny EMSG}}$ can be smaller or larger than the GR case $\dot{P}_{\text{GR}}$.  Here, we assume that the free parameter of the theory can take both positive and negative values. In \cite{akarsu2018constraint}, by studying the masses and radii of neutron stars, a bound on the EMSG parameter is fixed. It is shown that $-10^{-37}\,\text{m}\,\text{s}^2\,\text{kg}^{-1}<f_0'<+10^{-36}\,\text{m}\,\text{s}^2\,\text{kg}^{-1}$.
This limitation is also consistent with what is provided in \cite{nari2018compact}. 
Applying this bound, we examine the time derivative of the orbital period for the well-known Hulse-Taylor binary pulsar, J1915+1606, in EMSG.  To do so, we consider that the density of both the pulsar and the companion is of the order of the neutron star density $\sim 10^{17}\,\text{kg}\,\text{m}^{-3}$.  
In this case, we find that $-2.447\times 10^{-12}<\dot{P}_{\text{\tiny EMSG}}<-2.399\times10^{-12}$.  On the other hand, the observed orbital period variation of this system is of the order $\dot{P}_{\text{obs}}=-2.423\times10^{-12}$ \cite{hulse1975discovery,weisberg2010timing}. So, one can conclude that, at least this specific binary system does not rule out EMSG.

\begin{figure}
\begin{center}
\centering
\includegraphics[scale=0.65]{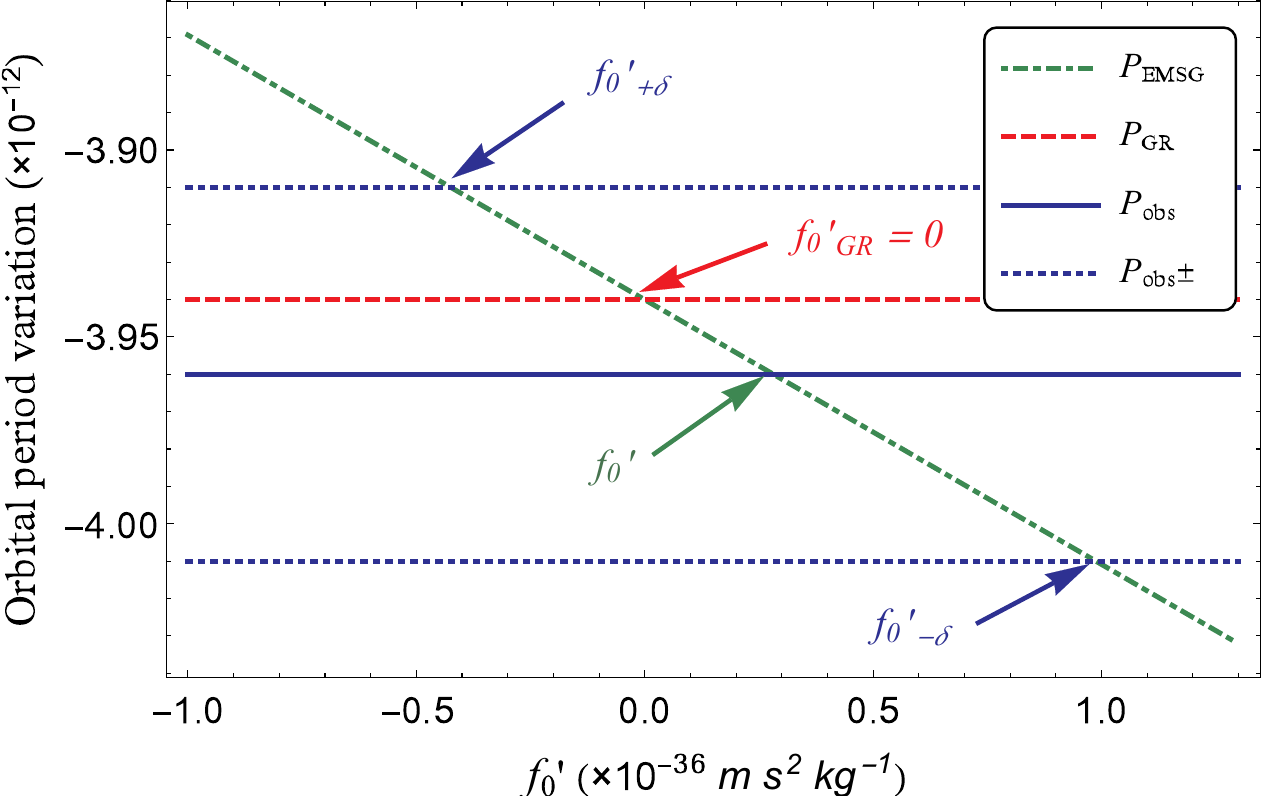}
\caption{The orbital period variation of the double neutron star binary pulsar  J2129+1210C. In this figure, the green dot-dashed line shows $\dot{P}_{\text{\tiny EMSG}}$ in terms of $f_0'$. cf. Eq. \eqref{Tdot}. We also insert $\dot{P}_{\text{GR}}$ and $\dot{P}_{\text{obs}}$ in this plot. Where the $\dot{P}_{\text{\tiny EMSG}}$ line intersects the blue solid and red dashed lines indicates $f_0'$ and ${f_0'}_{\text{GR}}$, respectively.
Moreover, the upper and lower limits of $f_0'$ are obtained from the intersection of the green dot-dashed and blue dotted lines. The numerical values of these are given in Table \ref{table2}.} \label{fig2}
\end{center}
\end{figure}

We apply the method described in \cite{de2013testing} to restrict the EMSG parameter. We consider that the EMSG orbital period variation could fully justify the observed one, i.e., we set $\dot{P}_{\text{\tiny EMSG}}=\dot{P}_{\text{obs}}$ or equivalently $\dot{P}_{\text{GR}}\big(1+2\alpha\big)=\dot{P}_{\text{obs}}$. So, the free parameter $f_0'$ is restricted as 
\begin{align}
\label{re_f0}
{f_0'}=\frac{1}{2c^2}\frac{m}{\big(\rho_1m_2+\rho_2m_1\big)}\Big[ \Big(\frac{\dot{P}_{\text{obs}}}{\dot{P}_{\text{GR}}}\Big)-1\Big].
\end{align}
Moreover, we assume that uncertainty on $f_0'$ is given by 
\begin{align}\label{limits_of_f0}
{f_0'}_{\pm \delta}=\frac{1}{2 c^2}\frac{m}{\big(\rho_1m_2+\rho_2m_1\big)}\Big[ \Big(\frac{\dot{P}_{\text{obs}}\pm \delta}{\dot{P}_{\text{GR}}}\Big)-1\Big],
\end{align}
where $\delta$ shows the experimental error on $\dot{P}_{\text{obs}}$. Here, we do not add the contributions resulting from the Galactic and kinematic/Shklovskii accelerations, $\dot{P}_{\text{Gal}}$ and $\dot{P}_{\text{kin}}$ \cite{damour1991orbital}, to the orbital decay. After these values are provided in the literature, the ratio $\dot{P}_{\text{obs}}/\dot{P}_{\text{GR}}$ in Eq. \eqref{re_f0} should be changed to $\big(\dot{P}_{\text{obs}}-\dot{P}_{\text{Gal}}-\dot{P}_{\text{kin}}\big)/\dot{P}_{\text{GR}}$.       
We select the binary pulsars whose characteristics like the mass of pulsar $m_{\text{p}}$ and mass of companion $m_{\text{c}}$, orbital eccentricity $e$, orbital period $P$, and observed orbital period deviation are introduced with a good precision in the literature.
In Table \ref{table1}, the J-name of these binary pulsars and their characteristics needed during our calculations are shown. 
For the first relativistic binary pulsar, J2129+1210C, we exhibit our method in Fig. \ref{fig2}.  As seen in this figure, the function $\dot{P}_{\text{\tiny EMSG}}$ intersects the horizontal line of $\dot{P}_{\text{obs}}$. This point is indicated with the green arrow. Therefore, for this value of $f_0'$, this theory can justify the observed orbital period variation. Here, we assume that the density of both neutron stars in this binary pulsar is of the order $10^{17}\,\text{kg}\,\text{m}^{-3}$. Moreover, the upper and lower limits of this parameter are obtained according to Eq. \eqref{limits_of_f0}. These are also displayed with the blue arrows. 
Furthermore, one can see that for the value $f_0'=0$, our result reduces to the GR case. We display this with ${f_0'}_{\text{GR}}=0$ in this figure.

A similar interpretation is applied for the rest of the binary pulsars. The results are summarized in Table \ref{table2}. 
In this table, for each system, we obtain a numerical value for $f_0'$ which satisfies the observed results.
For the double neutron star binary pulsars, we set $\rho_1=\rho_2=10^{17}\,\text{kg}\,\text{m}^{-3}$; and for the last two systems in this table, which are the pulsar-white-dwarf binaries, we consider that the density of pulsar and companion is of the order $10^{17}$ and $10^{9}\, \text{kg}\,\text{m}^{-3}$, respectively.
Also, we introduce the interval $\triangle f_0'$ as $\vert {f_0'}_{+\delta}-{f_0'}_{- \delta}\vert/2$. In fact, twice this interval shows where $\dot{P}_{\text{\tiny EMSG}}$ would intersect the observed case. 
 
The first three systems reveal that the $f_0'$ value of the order $10^{-37}\,\text{m}\,\text{s}^2\,\text{kg}^{-1}$ improves the theoretical predictions. In fact, in these cases, the same order of the EMSG parameter can fill the gap between the observed results and the GR predictions. 
On the other hand, the next double neutron star binary pulsar, J1537+1155, behaves differently. For this relativistic system, we find that ${f_0'}$ is negative and its value is different from that of the previous double neutron star binary pulsars. However, it is necessary to emphasize that for this system, the percent error $|(\dot{P}_{\text{obs}}-\dot{P}_{\text{GR}})/\dot{P}_{\text{GR}}|\times 100$ is about $28\%$, while the percent error of the previous systems is less than $1\%$, making it of less interest and reliability in this investigation.
Modifying Eq. \eqref{re_f0} by inserting $\dot{P}_{\text{Gal}}=(0.037\pm 0.011)\times 10^{-12}$ \cite{stairs2002studies} for J1537+1155 improves the results and increases ${f_0'}$ up to $-2.46\times10^{-36}\,\text{m}\,\text{s}^2\text{kg}^{-1}$. Moreover, we simply assume that the density of the components is constant with the order $10^{17}\,\text{kg}\,\text{m}^{-3}$. For the higher density range, ${f_0'}$ deduced from J1537+1155 may also enter the mentioned bound.

\begin{figure}
\centering
\includegraphics[scale=0.75]{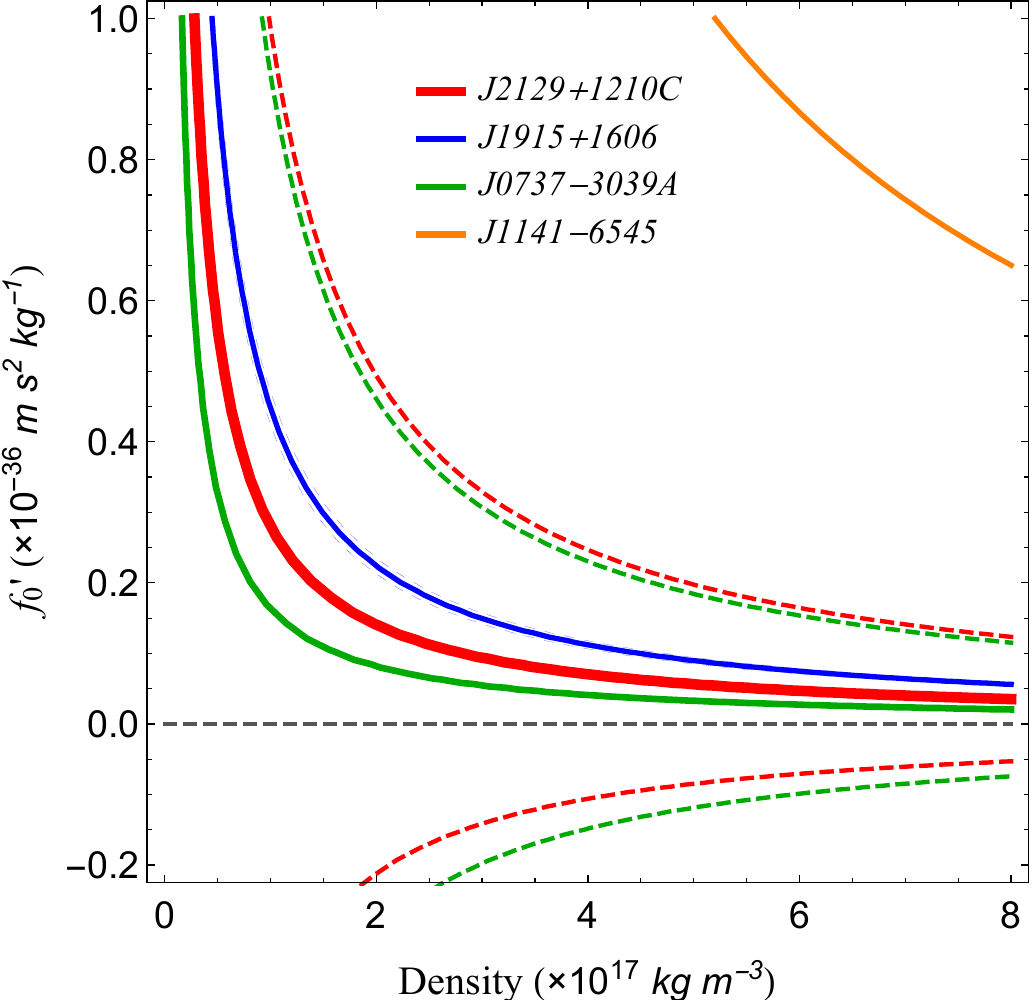}
\caption{The free parameter of the EMSG theory in terms of the density of the neutron star. Here, we consider that the density can change in the interval $(10^{17}-8\times10^{17})\,\text{kg}\,\text{m}^{-3}$. Here, ${f_0'}$ is displayed for the introduced relativistic binary pulsars. The colored solid and dashed curves represent the solutions   \eqref{re_f0} and \eqref{limits_of_f0}, respectively. 
Here, we show the range of $f_0'$ where most of the systems behave similarly. 
We find that the binary pulsars J1537+1155 and J1738+0333 have no solution in this range. Also, the dashed curves belonging to J1141-6545 do not enter this range.
}\label{fig3}
\end{figure}

For the last two pulsar-white-dwarf binaries, J1141-6545 and J1738+0333, we assume that the white dwarf density is of the order $10^{9}\,\text{kg}\,\text{m}^{-3}$. For these two cases, one can see that the value of ${f_0'}$ is not of the order of the first three cases. It can be a result of the effect of the Galactic and kinematic accelerations, the constant density assumption, or the magnitude of the density inside these systems.  To clarify the latter fact, we study Eq. \eqref{re_f0} for the different range of $f_0'$ and $\rho$ in Fig. \ref{fig3}. 
It should be mentioned in the case of white-dwarf-pulsar binaries, we assume that the density of the companion is of the order $10^{9}\,\text{kg}\,\text{m}^{-3}$, and we consider that the density of the pulsar can change between $10^{17}-8\times10^{17}\,\text{kg}\,\text{m}^{-3}$.
For Four binary pulsars J2129+1210C,  J1915+1606, J0737-3039A, and  J1141-6545, we obtain reliable values for $f_0'$ in terms of $\rho$. As seen, the value of $f_0'$ is completely sensitive to the density, and at the high-density regime, it decreases to $1$ order of magnitude. 
We also find that the binary pulsars J1537+1155 and J1738+0333 have no solution in the range shown in this figure. It should be emphasized that for these systems, the percent error $|(\dot{P}_{\text{obs}}-\dot{P}_{\text{GR}})/\dot{P}_{\text{GR}}|\times 100$ is more than $25\%$. Because of this high percent error, we ignore these two systems in the following analysis and focus our attention on the rest systems in Table \ref{table2} where the percent error is less than $5\%$.

       \begin{table}[!]
       \caption{The value of the EMSG parameter and its  uncertainty by analyzing the orbital decay of the specific binary pulsars. In SI units, the dimension of this parameter is $\text{m}\, \text{s}^2 \text{kg}^{-1}$. It should be mentioned, the best previous constraint on $f_0'$ is introduced in \cite{akarsu2018constraint} which is in the range $-10^{-37}\text{m}\,\text{s}^2\text{kg}^{-1}<f_0'<+10^{-36}\text{m}\,\text{s}^2\text{kg}^{-1}$.}\label{table2}
       \begin{center}
       \footnotesize
      \renewcommand{\arraystretch}{1.5}
         \begin{tabular}{lcccccccccccc}
         \hline         System  & && ${f_0'}$ & && ${f_0'}_{+\delta}$ &  && ${f_0'}_{-\delta}$ &  && $\triangle{f_0'}$\\
           && &  $(\times10^{-37})$ & && $(\times10^{-37})$ & &&  $(\times10^{-37})$ &&& $(\times10^{-37})$\\
         \hline
         \hline
        J2129+1210C  &&& 2.82 &&& $-4.24$ &&& 9.89 &&& 7.06\\
         J1915+1606 &&& 4.49 &&& 4.25 &&& 4.72 &&& 0.23 \\
         J0737-3039A &&& 1.64 &&& $-5.94$ &&& 9.23 &&& 7.58  \\
         J1537+1155 &&& $-157.54$ &&& $-157.83$ &&& $-157.25$ &&& 0.29\\
         J1141-6545 &&& 52.02 &&& $-28.69$ &&& 132.73 &&& 80.71 \\
         J1738+0333 &&& $-1922.75$ &&& $-2491.89$ &&& $-1353.62$ &&& 569.13 \\ 
        \hline
         \end{tabular}
       \end{center} 
       \normalsize
       \end{table}

\section{Conclusion}\label{Conclusion}

\begin{figure*}
\centering
\includegraphics[scale=0.68]{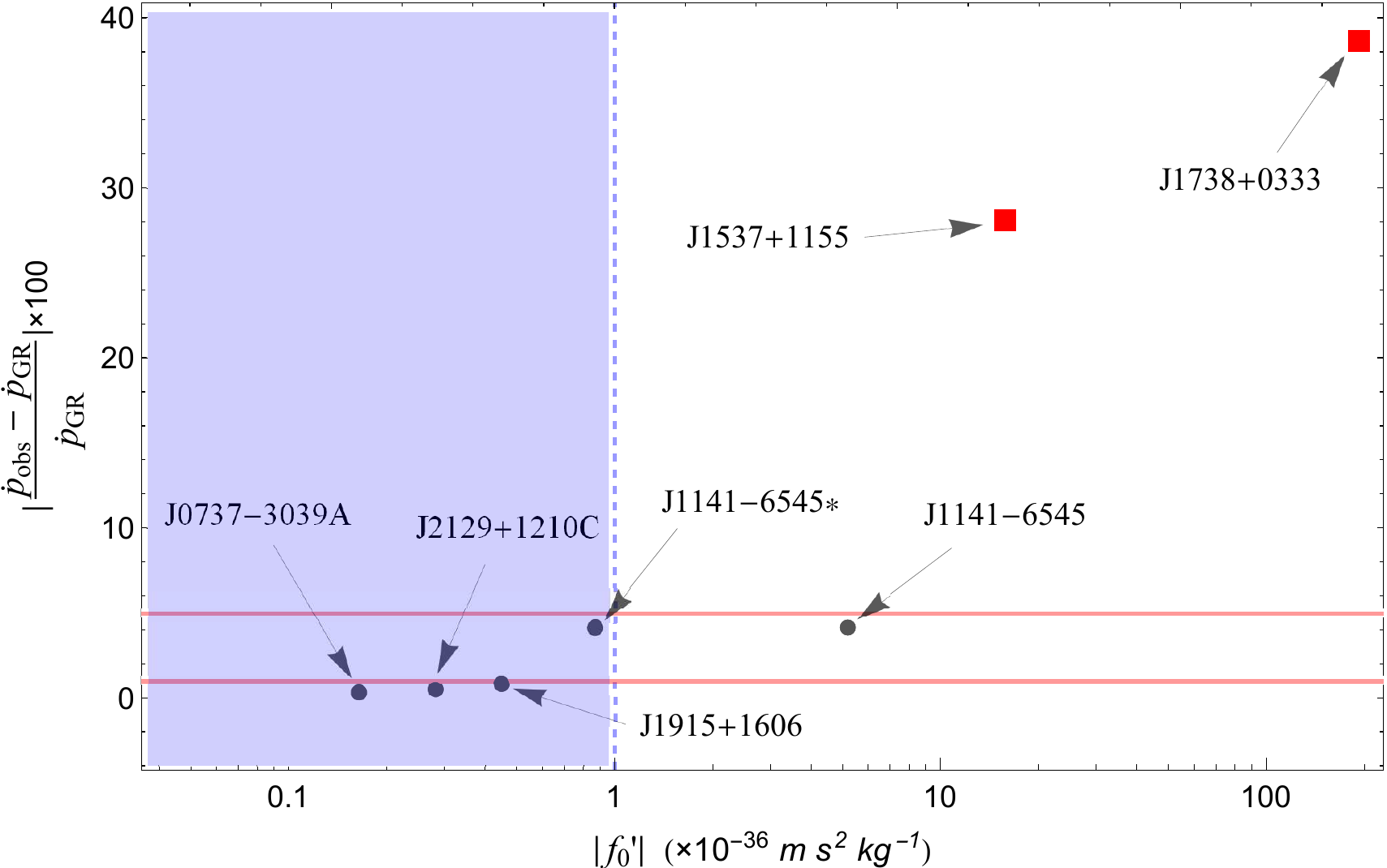}
\caption{The absolute value of the free parameter of the EMSG theory in terms of the percent error $|(\dot{P}_{\text{obs}}-\dot{P}_{\text{GR}})/\dot{P}_{\text{GR}}|\times 100$.
For the sake of convenience, the horizontal axis is scaled logarithmically. 
Here, except for $\text{J}1141\text{-}6545*$, we consider that the density of the neutron star is of the order $10^{17}\text{kg}\,\text{m}^{-3}$. For $\text{J}1141\text{-}6545*$, we assume that the density of the pulsar is of the order $6\times10^{17}\text{kg}\,\text{m}^{-3}$. The horizontal lines, from top to bottom, represent the percent error $5\%$ and $1\%$, respectively. Moreover, the colored area shows the allowed region for $f_0'$ introduced in \cite{akarsu2018constraint}. Four cases J0737-3039A, J2129+1210C, J1915+1606, and $\text{J}1141\text{-}6545*$ are well located in this region. These systems are displayed by the black circle. The rest systems where the percent error is more than $5\%$ are indicated by the red square. }\label{fig4}
\end{figure*}

In this paper, we have introduced the PM approximation of EMSG. Utilizing this approximation, the gravitational energy flux has been studied in this theory. As an application of our derivation, we have investigated the EMSG effect on the post-Keplerian parameter $\dot{P}$. Comparing $\dot{P}_{\text{EMSG}}$ with $\dot{P}_{\text{obs}}$ of six binary pulsars categorized in Table \ref{table1}, the free parameter of the EMSG theory has been fixed. The results are summarized in Table \ref{table2}.

It has been shown that for the first three binary pulsars in this table, the parameter $f_0'$ is of the order of $10^{-37}\,\text{m}\,\text{s}^2\text{kg}^{-1}$. In other words, for this value of the EMSG parameter, the gap between the observed results and the GR predictions can be filled. However, the order of magnitude of $f_0'$ is quite different for the next three binary pulsars. This discrepancy is visualized in Fig. \ref{fig4}. In this figure, the absolute value of $f_0'$s in terms of $|(\dot{P}_{\text{obs}}-\dot{P}_{\text{GR}})/\dot{P}_{\text{GR}}|\times 100$ for the six binary pulsars is displayed.

Some assumptions taken during our derivation can be the origin of this discrepancy. As mentioned, we have considered that the density of the binary components is constant, and for the neutron star, we set $\rho=10^{17} \,\text{kg}\,\text{m}^{-3}$. On the other hand, determining the EMSG parameter strongly depends on the density of the subject system. We have shown that increasing the density to $8\times10^{17} \,\text{kg}\,\text{m}^{-3}$, the value of $f_0'$ can change about 1 order of magnitude. In Figs. \ref{fig3} and \ref{fig4}, it is exhibited that at the higher density range, the binary pulsar J1141-6545 may also behave like the first three so that $f_0'$ enters the range $<10^{-36} \,\text{m}\,\text{s}^2\text{kg}^{-1}$. This case is displayed by an asterisk in the J-name of this binary pulsar in Fig. \ref{fig4}. 

In addition, another reason for this discrepancy may be the mass dependence of $f_0'$. In Eq. \eqref{re_f0}, we have illustrated that the mass-dependent terms exist in the definition of $f_0'$. Notice that some part of this dependency lies in $\dot{P}_{\text{GR}}$. So, in an accurate analysis to determine $f_0'$, one needs to study the mass of the binary components in the EMSG framework. To do so, one should obtain the other post-Keplerian parameters in EMSG. In this paper, as the first step to estimate the order of the free parameter in this theory by studying $\dot{P}$, we have utilized the masses derived in the GR context. We leave the complete analysis for future work.

More importantly, other causes may come from the considerable contributions of the Galactic and kinematic/Shklovskii accelerations to the orbital decay \cite{damour1991orbital}. In this case, the ratio $\dot{P}_{\text{obs}}/\dot{P}_{\text{EMSG}}$ can dramatically change and affect the value of $f_0'$. We have argued that by adding the Galactic acceleration, the value of $f_0'$ obtained from the system J1537+1155 is improved so that it approaches that of the first three systems.   
So, to estimate the free parameter of EMSG truly, besides obtaining the mass and density of the binary components more accurately, we need to consider the role of the Galactic and kinematic/Shklovskii accelerations. So, to estimate the  value of $f_0'$, we ignore systems with high percent error and focus our attention on the rest systems where the percent error is less than $5\%$.

\begin{figure}
\begin{center}
\centering
\includegraphics[scale=0.68]{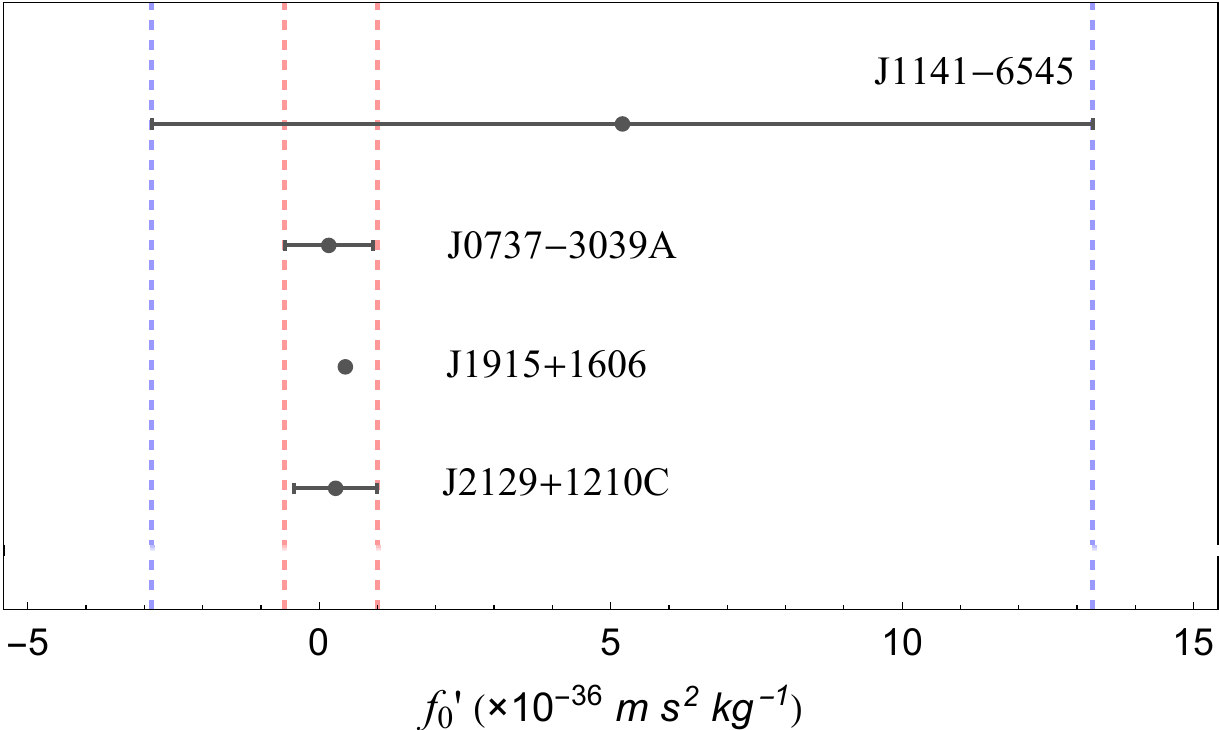}
\caption{ The value of $f_0'$ and its error bar for the systems with percent error $<5\%$.
The dashed lines show the estimated bound for $f_0'$. Here, we choose the worse error to estimate this bound. The blue dashed lines represent the region deduced from the four systems and the red ones exhibit the bound found from the binary pulsars J2129+1210C,  J1915+1606, and J0737-3039A.} \label{fig5}
\end{center}
\end{figure}

Figure. \ref{fig5} represents the value of $f_0'$ and its error bar for the systems with percent error $<5\%$.  As seen from this figure, for the binary pulsars J2129+1210C,  J1915+1606, and J0737-3039A  with the percent error less than $1\%$,  for $f_0'<10^{-36}$, one can fill the gap between the observed results and the GR predictions. Considering the error bars, we conclude  that $-6\times 10^{-37}\text{m}\,\text{s}^2\text{kg}^{-1}<f_0'<+10^{-36}\text{m}\,\text{s}^2\text{kg}^{-1}$ from these binary pulsars. This constraint on $f_0'$ is in agreement with the specified range $-10^{-37}\text{m}\,\text{s}^2\text{kg}^{-1}<f_0'<+10^{-36}\text{m}\,\text{s}^2\text{kg}^{-1}$ introduced in \cite{akarsu2018constraint}. On the other hand, adding J1141-6545, one can estimate $-2.8\times 10^{-36}\text{m}\,\text{s}^2\text{kg}^{-1}\leq f_0' \leq +1.3\times 10^{-35}\text{m}\,\text{s}^2\text{kg}^{-1}$ for the range of the free parameter of theory. Although, because of the different nature of this white-dwarf-pulsar binary from the first three systems, more effects like the mass transfer in the binary system and the accurate density of the components should be considered.  
In conclusion, the first three binary pulsars in Table \ref{table2} do not rule out the theory, and by analyzing these system, we estimate that the bound $-6\times 10^{-37}\text{m}\,\text{s}^2\text{kg}^{-1}<f_0'<+10^{-36}\text{m}\,\text{s}^2\text{kg}^{-1}$ for the free parameter of EMSG.
However, for the last three binaries where the percent error is high, further study on the density, the mass of the binary pulsar components, (in the case of white-dwarf-pulsar binaries) the hydrodynamics effects due to the mass transfer, and the contributions of the Galactic and kinematic/Shklovskii accelerations to the orbital decay are required to improve the free parameter estimation.

\section*{ACKNOWLEDGMENTS}

We thank the anonymous referees for their useful and constructive comments and for introducing related references.
This work is supported by Ferdowsi University of Mashhad under Grant No. 1/53355 (09/10/1399). I.D.M acknowledges support from MICINN (Spain) under de project IJCI2018-036198-I. I.D.M is also supported by Junta de Castilla y León (SA096P20), and Spanish Ministerio de Ciencia, Innovación y Universidades and FEDER (PGC2018-096038-B-I00).

\bibliographystyle{apsrev4}
\bibliography{short,GWs_in_EMSG}

\clearpage
\appendix

\section{Symbols and Acronyms}

In this appendix, for the convenience of the reader, symbols and acronyms are summarized in Tables. \ref{table4} and \ref{table3}.

       \begin{table}[!]
       \caption{List of abbreviations frequently used in the paper.}\label{table4}
       \begin{center}
       \footnotesize
      \renewcommand{\arraystretch}{1.5}
         \begin{tabular}{lllll }
         \hline         Acronyms  &&& Extended name & \\
         \hline
         \hline
         GR  &&& General theory of relativity & \\
         GW  &&&  Gravitational wave &  \\
         EMSG &&& Energy-momentum-squared gravity & \\
         PM &&& Post-Minkowskian  & \\
         T-T &&& Transverse-tracefree &\\
         PN &&& Post-Newtonian &\\ 
        $\text{cm}_A$ &&& Center of mass of the body $A$ &\\
         SEP &&& Strong equivalence principle & \\
        \hline
         \end{tabular}
       \end{center} 
       \normalsize
       \end{table}

       \begin{table}[!]
       \caption{List of symbols used in the paper.}\label{table3}
       \begin{center}
       \footnotesize
      \renewcommand{\arraystretch}{1.5}
         \begin{tabular}{lllll }
         \hline         Symbols  &&& Explanation & \\
         \hline
         \hline
        ${\boldsymbol{T}}^2$  &&& $T^{\alpha\beta}T_{\alpha\beta}$ & \\
         $T_{\text{eff}}^{\mu\nu} $ &&&  Effective energy-momentum tensor & \\
         $\mathfrak{g}^{\mu\nu}$ &&&  Gothic metric defined by $\sqrt{-g}g^{\mu\nu}$  &\\
         $h^{\mu\nu}$ &&& Gravitational potential &\\
        $\tau^{\mu\nu}_{\text{eff}} $ &&& Effective energy-momentum pseudotensor &\\ 
        $(-g)t_{\text{LL}}^{\mu\nu}$ &&&  Landau-Lifshitz pseudotensor &\\
        $(-g)t_{\text{H}}^{\mu\nu}$ &&&  Harmonic pseudotensor &\\
        $P^{\alpha} $ &&&  Total four-vector momentum &\\
        $f'_0$ &&& Free parameter of the EMSG theory &  \\
        $\mathcal{P}$ &&& Flux of energy &\\
         $P$ &&& Orbital period of binary system& \\
       $\lambda_c$ &&& Characteristic wavelength of the GW signals &\\
       $\tau$ &&& Retarded time &\\
       $\rho$ &&& Mass density & \\
       $\rho^*$ &&& Rescaled mass density & \\
        $\epsilon$ &&& Proper internal energy density &\\
        $p$ &&& Pressure &\\
        $\Pi$ &&&  Fluid's internal energy per unit mass &\\
        $c$ &&& Speed of light& \\
        $U$ &&& Newtonian  potential &\\
        $U_{\text{\tiny EMS}}$ &&& EMSG potential defined in Eq. \eqref{U_EMSG} &\\
        $\mathcal{M}$ &&& Three-dimension sphere separating & \\
      &&&  the near zone from the wave zone &\\
     $\mathcal{R}$ &&&  Radius of $\mathcal{M}$ &\\
     $\partial\mathcal{M}$ &&& Boundary of $\mathcal{M}$ &\\
     $\mathcal{N}$ &&& Near zone &\\
     $\mathcal{W}$ &&& Wave zone &\\
     $ \mathfrak{M}$ &&& New quantity in EMSG defined in Eq. \eqref{D1}&\\
     $M_0$ &&&  Total matter inside the near zone &\\
     $\mathfrak{P}^j$ &&& New quantity in EMSG defined in Eq. \eqref{new} &\\
     $P_0^j$ &&& Near-zone momentum&\\
     $M$ &&& Chirp mass&\\
      $l$ &&& Semilatus rectum of the orbit & \\
    $e$ &&&  Orbital eccentricity &\\
        \hline
         \end{tabular}
       \end{center} 
       \normalsize
       \end{table}

\section{Essential Relations}\label{app_1}

In this Appendix, we introduce the essential relations required to study the balance-energy equation in the context of the EMSG theory. 
The first one is the  definition of the Landau-Lifshitz pseudotensor $(-g)t_{\text{LL}}^{\alpha\beta}$. Its general definition in terms of the gothic metric $\mathfrak{g}^{\alpha\beta}$ is given by
\begin{widetext}
\begin{align}\label{tLL}
\nonumber
(-g)t_{\text{LL}}^{\alpha\beta}=&\frac{c^4}{16\pi G}\bigg\lbrace\partial_{\lambda}\mathfrak{g}^{\alpha\beta}\partial_{\mu}\mathfrak{g}^{\lambda\mu}-\partial_{\lambda}\mathfrak{g}^{\alpha\lambda}\partial_{\mu}\mathfrak{g}^{\beta\mu}+\frac{1}{2}g^{\alpha\beta}g_{\lambda\mu}\partial_{\rho}\mathfrak{g}^{\lambda\nu}\partial_{\nu}\mathfrak{g}^{\mu\rho}-g^{\alpha\lambda}g_{\mu\nu}\partial_{\rho}\mathfrak{g}^{\beta\nu}\partial_{\lambda}\mathfrak{g}^{\mu\rho}-g^{\beta\lambda}g_{\mu\nu}\partial_{\rho}\mathfrak{g}^{\alpha\nu}\partial_{\lambda}\mathfrak{g}^{\mu\rho}\\
&+g_{\lambda\mu}g^{\nu\rho}\partial_{\nu}\mathfrak{g}^{\alpha\lambda}\partial_{\rho}\mathfrak{g}^{\beta\mu}+\frac{1}{8}\big(2g^{\alpha\lambda}g^{\beta\mu}-g^{\alpha\beta}g^{\lambda\mu}\big)\big(2g_{\nu\rho}g_{\sigma\tau}-g_{\rho\sigma}g_{\nu\tau}\big)\partial_{\lambda}\mathfrak{g}^{\nu\tau}\partial_{\mu}\mathfrak{g}^{\rho\sigma}\bigg\rbrace.
\end{align}
\end{widetext}
By defining the gravitational potential $h^{\alpha\beta}=\eta^{\alpha\beta}-\mathfrak{g}^{\alpha\beta}$ and imposing the harmonic gauge condition $\partial_{\beta}h^{\alpha\beta}=0$, one can simplify the above definition as
\begin{align}\label{tLL_harmonic}
\nonumber
&(-g)t_{\text{LL}}^{\alpha\beta}=\frac{c^4}{16\pi G}\bigg\lbrace\frac{1}{2}\eta^{\alpha\beta}\eta_{\lambda\mu}\partial_{\rho}h^{\lambda\nu}\partial_{\nu}h^{\mu\rho}-\eta^{\alpha\lambda}\eta_{\mu\nu}\partial_{\rho}h^{\beta\nu}\\\nonumber
&\times\partial_{\lambda}h^{\mu\rho}-\eta^{\beta\lambda}\eta_{\mu\nu}\partial_{\rho}h^{\alpha\nu}\partial_{\lambda}h^{\mu\rho}+\eta_{\lambda\mu}\eta^{\nu\rho}\partial_{\nu}h^{\alpha\lambda}\partial_{\rho}h^{\beta\mu}\\
&+\frac{1}{8}\big(2\eta^{\alpha\lambda}\eta^{\beta\mu}-\eta^{\alpha\beta}\eta^{\lambda\mu}\big)\big(2\eta_{\nu\rho}\eta_{\sigma\tau}-\eta_{\rho\sigma}\eta_{\nu\tau}\big)\partial_{\lambda}h^{\nu\tau}\partial_{\mu}h^{\rho\sigma}\bigg\rbrace.
\end{align}

The second set of the important relations we need during our calculation is the PM expansion of the metric and its determinant in terms of the gravitational potentials. In \cite{poisson2014gravity}, by utilizing the Landau-Lifshitz formalism and introducing the modern approach to the PM theory, it is shown that these PM expansions are given by the following general forms:
\begin{align}
\nonumber
& g_{\alpha\beta}=\eta_{\alpha\beta}+h_{\alpha\beta}-\frac{1}{2}h\eta_{\alpha\beta}+h_{\alpha\mu}h^{\mu}_{\beta}-\frac{1}{2}h h_{\alpha\beta}\\\label{PM_metric}
&~~~~~~+\big(\frac{1}{8}h^2-\frac{1}{4}h^{\mu\nu}h_{\mu\nu}\big)\eta_{\alpha\beta}+O(G^3),\\
\label{PM_g}
& (-g)=1-h+\frac{1}{2}h^2-\frac{1}{2}h^{\mu\nu}h_{\mu\nu}+O(G^3),
\end{align}
in which $h=\eta_{\alpha\beta}h^{\alpha\beta}=-h^{00}+h^{kk}$. It should be noted that to derive these relations, according to the PM expansion of $h_{\alpha\beta}$, it is considered that the gravitational potentials are at least of the order $O(G)$.
We use this general relation to find the near-zone metric in Sec. \ref{PM_expansion_of_metric}.

The third set of the essential relations is the solutions of the wave equation \eqref{wave_eq}. All general forms of these solutions with respect to the position of the field and source points are completely derived and classified in \cite{poisson2014gravity}. Here, we rewrite the solutions that we use in our calculation. 
The general solution of the wave equation when the source and field points both are situated within the near zone is given by 
\begin{align}
 \nonumber
 h_{\mathcal{N}}^{\alpha\beta}{(t,\boldsymbol{x})}=&\frac{4G}{c^4}\sum_{l=0}^{\infty}\frac{(-1)^l}{l! c^l}\\\label{hNear}
&\times\Big(\frac{\partial}{\partial t}\Big)^l
\int_{\mathcal{M}}\tau^{\alpha\beta}_{\text{eff}}{(t,\boldsymbol{x}')}\rvert{\boldsymbol{x}-\boldsymbol{x}'}\rvert^{l-1} d^3x'
\end{align}
Here, $\mathcal{M}$ represents the three-dimension region which in fact separates the near zone from the wave zone. The length scale and surface of this boundary are displayed by $\mathcal{R}$ and $\partial\mathcal{M}$, respectively.  
We should recall that in the primary stage of the iterated method, we are not allowed to impose the harmonic gauge conditions, $\partial_\alpha h^{\alpha\beta}=0$, or equivalently the conservation equations, $\partial_\alpha\tau^{\alpha\beta}_{\text{eff}}=0$. In fact, we postpone this condition until the last step of the iterative procedure,  cf. \cite{poisson2014gravity}.
So, in our calculation, if we are in the primary stage of the iterated method,  we use Eq. \eqref{hNear} to find the near-zone portion of the potential in the near zone; and if we are in the very last iterated step, we can freely use $\partial_\alpha\tau^{\alpha\beta}_{\text{eff}}=0$ to simplify this integral and find the potential.

The next solution we need to complete our derivation is the near-zone solution of the wave equation where the field point is in the wave zone and the source point is situated in the near zone. In this case, unlike the previous near-zone solution, the field point is located
in the wave zone. 
The general form of this solution is given by
 \begin{align}\label{hNwave}
h_{\mathcal{N}}^{\alpha\beta}(t,\bm{x})=&\frac{4G}{c^4}\sum_{l=0}^{\infty}\frac{(-1)^{l}}{l!}
 \\\nonumber
 &\times\partial_{j_1j_2\cdots j_l}\bigg[\frac{1}{r}\int_{\mathcal{M}}\tau^{\alpha\beta}_{\text{eff}}(\tau,\bm{x}')x'^{j_1j_2\cdots j_l}d^3x'\bigg].
 \end{align}
Here $x^{j_1j_2\cdots j_l}$ stands for $x^{j_1}x^{j_2}\cdots x^{j_l}$ and $\partial_{j_1j_2\cdots j_l}$ shows $\partial_{j_1}\partial_{j_2}\cdots \partial_{j_l}$. We recall that $\tau=t-r/c$ is the retarded time and the integrand in this integral unlike the previous solution is a function of $\tau$.

Another solution that we will encounter during our calculation, is the wave-zone solution  where the field and source points both are situated in the wave zone. In this case, considering the source of the wave-zone portion of the gravitational potential as 
\begin{align}\label{eq7}
\tau^{\alpha\beta}_{\text{\tiny EMS}}=\frac{1}{4\pi}\frac{f^{\alpha\beta}(\tau)}{r^n}n^{<j_1j_2\cdots j_l>},
\end{align}
the wave-zone portion of the gravitational potential is given in the following form 
\begin{align}\label{eq8}
\nonumber
 h^{\alpha\beta}_{\mathcal{W}}(t,\bm{x})&=\frac{4 G}{c^4}\frac{n^{<j_1j_2\cdots j_l>}}{r}\bigg\lbrace\int_{0}^{\mathcal{R}}f^{\alpha\beta}(\tau-2s/c)A(s,r)ds\\
&+\int_{\mathcal{R}}^{\infty}f^{\alpha\beta}(\tau-2s/c)B(s,r)ds\bigg\rbrace,
\end{align}
in which $A(s,r)=\int_{\mathcal{R}}^{r+s}P_{l}(\zeta)p^{1-n}dp$ and $B(s,r)=\int_{s}^{r+s}P_{l}(\zeta)p^{1-n}dp$. Here, $P_{l}(\zeta)$ is a Legendre polynomial, $\zeta=(r+2s)/r-2s(r+s)/(rp)$, and $n^{<j_1j_2\cdots j_l>}$ is an angular symmetric trace-free tensor introduced in Eq. (1.154) of \cite{poisson2014gravity}.

\section{On the time derivative of $\mathfrak{M}$ }\label{app_2}

During our calculations in Sec. \ref{PM approximation in EMSG}, we have encountered the new EMSG parameter $\mathfrak{M}$ introduced by Eq. \eqref{D1}. We have then shown that the time derivative of this parameter is entered in several PN corrections. Here, we try to simplify $d \mathfrak{M}/dt$.
Given Eq. \eqref{D1}, we have
\begin{align}
\frac{d \mathfrak{M}}{dt}=\int_{\mathcal{M}}\frac{\partial}{\partial t}{\rho^*}^2(t,\boldsymbol{x})d^3x.
\end{align}
Utilizing the conservation of the rest-mass density \eqref{rho}, we find that
\begin{align}
\frac{d \mathfrak{M}}{dt}=-2\int_{\mathcal{M}}\rho^*\partial_j\big(\rho^*v^j\big)d^3x.
\end{align}
One can simply show that this relation can be written as
\begin{align}
\frac{d \mathfrak{M}}{dt}=-2\oint_{\partial \mathcal{M}}{\rho^*}^2v^jdS_j+2\int_{\mathcal{M}}\partial_j \rho^*\big(\rho^*v^j\big)d^3x,
\end{align}
after applying the Gauss divergence theorem. Noting that the matter part vanishes on the boundary of $\mathcal{M}$, the first surface integral is zero. So, this term is reduced to 
\begin{align}
\frac{d \mathfrak{M}}{dt}=2\int_{\mathcal{M}}\rho^*v^j\partial_j \rho^*d^3x
\end{align}
As seen, $d \mathfrak{M}/dt$ is finally proportional to the integration of $\rho^* v^j\partial_j \rho^*$ and does not necessarily vanish.

\section{N-body description of the gravitational potentials}\label{app_3}

As we have seen in Sec. \ref{Gravitational potentials in the second iteration}, it is important to know the order of the magnitude of the Newtonian and EMSG potentials $U$ and $U_{\text{\tiny EMS}}$ on the boundary between the near and wave zones. 
We have called it $\partial\mathcal{M}$.
For this purpose, in this Appendix, we benefit from the N-body description of the fluid system to derive an appropriate relation which can approximately define these gravitational potentials on the surface $\partial\mathcal{M}$. 
 
To do so, we consider that the fluid is constructed from $N$ well-separated bodies. In this description, we can rewrite the gravitational potential as
\begin{align}
U(t,\bm{x})=\sum_{A=1}^{N}G\int_{A}\frac{\rho^*(t,\bm{x}')}{\vert \bm{x}-\bm{x}'\vert}d^3x',
\end{align}
where the index ``$A$" stands for the body $A$ and $A=1, 2, ..., N$. We assume that the body $A$ occupies the volume $V_A$ in space with length scale $R_A$.  
To simplify this integral, we change the variable as $\bm{x}'=\bm{\bar{x}}'+\bm{r}_A$. Here, $\bm{r}_A(t)$ is the position of the center of mass of the body $A$ and the vector $\bm{\bar{x}}$ is the distance between a fluid element and $\bm{r}_A(t)$. So, we have
\begin{align}\label{U(t,x)}
U(t,\bm{x})=\sum_{A=1}^{N}G\int_{A}\frac{\rho^*(t,\bm{r}_A+\bm{\bar{x}}')}{\vert \bm{s}_A-\bm{\bar{x}}'\vert}d^3x',
\end{align}
in which $s_A=\vert \bm{x}-\bm{r}_A \vert$ is the position of the field point relative to the center of mass of the body $A$.
Now, we assume that this distance is much larger than the dimension of the body $A$, i.e., we have $R_A\ll s_A$.
Using this condition, we can expand $\vert \bm{s}_A-\bm{\bar{x}}'\vert^{-1}$ in powers of $\bm{\bar{x}}'$. Regarding this fact, one can simplify Eq. \eqref{U(t,x)} to
\begin{align}\label{U}
U=\sum_{A=1}^{N}\Big(\frac{G m_A}{s_A}+\frac{1}{2}I_A^{jk}\partial_{jk}\frac{1}{s_A}+\cdots\Big),
\end{align}
where 
\begin{subequations}
\begin{align}
&m_A=\int_A{\rho^*}'d^3\bar{x}',\\
&I_A^{jk}=\int_A{\rho^*}'\bar{x}'^j\bar{x}'^kd^3\bar{x}',
\end{align}
\end{subequations}
are the material mass and quadrupole moment of the body $A$, respectively. This derivation is comprehensively introduced in chapter 9 of \cite{poisson2014gravity}.
Now, we assume that each compact body is located in the near zone, i.e., $r_A\ll \mathcal{R}$. 
Bearing this assumption in mind, one can deduce that Eq. \eqref{U} reduces to
\begin{align}\label{U_N-body}
U=\sum_{A}\frac{G m_A}{r}+O(r^{-3}),
\end{align}
when we want to study the gravitational potential in the wave zone, i.e., $\vert \bm{x} \vert \geq\mathcal{R}$. Here, we set $s_A\simeq\vert \bm{x }\vert=r$. 
In Sec. \ref{Gravitational potentials in the second iteration}, we use this potential as a source of the surface integrals. We then set $r =\mathcal{R}$ in that derivation.

In order to obtain the EMSG potential, the other source term of the surface integrals, we apply the same strategy introduced above.
In the N-body description, for Eq. \eqref{U_EMSG}, we have  
\begin{align}
U_{\text{\tiny EMS}}(t,\bm{x})=\sum_{A}G\int_{A}\frac{{\rho^*}^2(t,\bm{x}')}{\vert \bm{x}-\bm{x}'\vert}d^3x'.
\end{align} 
Changing the integration variable like before and using the Taylor expansion of $\vert \bm{s}_A-\bm{\bar{x}}'\vert^{-1}$, we arrive at 
\begin{align}
U_{\text{\tiny EMS}}=\sum_{A}\Big(\frac{G \mathfrak{M}_A}{s_A}+\frac{1}{2}\mathfrak{I}_A^{jk}\partial_{jk}\frac{1}{s_A}+\cdots\Big),
\end{align}
where
\begin{subequations}
\begin{align}
\label{M_A}
& \mathfrak{M}_{A}=\int_A{\rho^*}'^2d^3\bar{x}',\\
\label{I_A^jk}
&\mathfrak{I}_A^{jk}=\int_A{\rho^*}'^2\bar{x}'^j\bar{x}'^kd^3\bar{x}'.
\end{align}
\end{subequations}
It should be mentioned that to simplify this relation, we consider the following symmetry for the matter density $\rho^*(t,\bm{r}_A-\bm{\bar{x}})=\rho^*(t,\bm{r}_A+\bm{\bar{x}})$ and deduce that the integral $\int_A {\rho^*}'^2\bar{x}'^jd^3\bar{x}'$ vanishes. The same scheme is also applied in the previous case. One can easily show that this relation is simplified to
\begin{align}\label{U_EMSG_N-body}
U_{\text{\tiny EMS}}=\sum_{A}\frac{G \mathfrak{M}_{A}}{r}+O(r^{-3}),
\end{align}
when $\vert \bm{x}\vert\gg\vert\bm{r}_A\vert$.

\end{document}